\documentclass{aa}  

\usepackage{graphicx}
\usepackage{color}
\usepackage{todonotes}
\usepackage[normalem]{ulem}
\usepackage{amstext}
\usepackage{txfonts}
\definecolor{dkgreen}{rgb}{0,0.6,0}
\definecolor{gray}{rgb}{0.5,0.5,0.5}
\definecolor{mauve}{rgb}{0.58,0,0.82}
\usepackage{listings}
 \usepackage[breaklinks=true]{hyperref}

\newcommand{\webref}[2]{\href{#1}{\color{blue}#2}}

\newcommand{\Gaia}{\textit{Gaia}}

\newcommand{\GDR}{\Gaia\ DR}
\newcommand{\EDR}{\Gaia\ EDR}
\newcommand{\G}{$G$}
\newcommand{\BP}{$G_{\rm BP}$}
\newcommand{\RP}{$G_{\rm RP}$}
\newcommand{\bprp}{\ensuremath{G_{\rm BP}-G_{\rm RP}}\xspace}
\newcommand{\orcit}[1]{\protect\href{https://orcid.org/#1}{\protect\includegraphics[width=8pt]{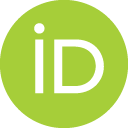}}}
\def\ie{{\sl i.e.}\xspace}
\def\eg{{\sl e.g.}\xspace}

\newcommand{\cxs}{\ensuremath{C^{\ast}}\xspace}

\begin{document}

   \title{Gaia Data Release 3: The \Gaia\ Andromeda Photometric Survey}

   \author{D.W. Evans\orcit{0000-0002-6685-5998} 
        \inst{\ref{inst:ioa}}
        \and
        L. Eyer\orcit{0000-0002-0182-8040} 
        \inst{\ref{inst:geneva}}
        \and
        G.~Busso\orcit{0000-0003-0937-9849} 
        \inst{\ref{inst:ioa}}
        \and
        M. Riello\orcit{0000-0002-3134-0935} 
        \inst{\ref{inst:ioa}}
        \and
        F.~De Angeli\orcit{0000-0003-1879-0488} 
        \inst{\ref{inst:ioa}}
        \and
        P.W.~Burgess 
        \inst{\ref{inst:ioa}}
        \and        
        M.~Audard\orcit{0000-0003-4721-034X} 
        \inst{\ref{inst:geneva},\ref{inst:ecogia}}
        \and
        G.~Clementini\orcit{0000-0001-9206-9723} 
        \inst{\ref{inst:bol}}
        \and
        A.~Garofalo\orcit{0000-0002-5907-0375} 
        \inst{\ref{inst:bol}}
        \and 
        B.~Holl\orcit{0000-0001-6220-3266} 
        \inst{\ref{inst:geneva},\ref{inst:ecogia}}
        \and
        G.~Jevardat de Fombelle
        \inst{\ref{inst:geneva}}
        \and
        A.C.~Lanzafame\orcit{0000-0002-2697-3607} 
        \inst{\ref{inst:ucatania},\ref{inst:catania}}
        \and 
        I.~Lecoeur-Taibi\orcit{0000-0003-0029-8575} 
        \inst{\ref{inst:ecogia}}
        \and
        N.~Mowlavi\orcit{0000-0003-1578-6993} 
        \inst{\ref{inst:geneva}}
        \and
        \inst{\ref{inst:ecogia}}
        K.~Nienartowicz\orcit{0000-0001-5415-0547} 
        \inst{\ref{inst:sednai},\ref{inst:ecogia}}
        \and
        L.~Palaversa\orcit{0000-0003-3710-0331} 
        \inst{\ref{inst:zagreb}}
        \and
        L.~Rimoldini\orcit{0000-0002-0306-585X} 
        \inst{\ref{inst:ecogia}}
 }

   \institute{
        Institute of Astronomy, University of Cambridge, Madingley Road, Cambridge CB3 0HA,UK\\
        \email{dwe@ast.cam.ac.uk}
        \label{inst:ioa}
         \and
        Department of Astronomy, University of Geneva, Chemin Pegasi 51, CH-1290 Versoix, Switzerland
        \label{inst:geneva}
        \and 
        INAF - Osservatorio di Astrofisica e Scienza dello Spazio di Bologna, via Piero Gobetti 93/3, 40129 Bologna, Italy
        \label{inst:bol}
        \and 
        Department of Astronomy, University of Geneva, Chemin d'Ecogia 16, 1290 Versoix, Switzerland\label{inst:ecogia}
        \and 
        Dipartimento di Fisica e Astronomia ""Ettore Majorana"", Universit\`{a} di Catania, Via S. Sofia 64, 95123 Catania, Italy\label{inst:ucatania}
        \and 
        INAF - Osservatorio Astrofisico di Catania, via S. Sofia 78, 95123 Catania, Italy\label{inst:catania}\vfill
        \and 
        Sednai S\`{a}rl, Geneva, Switzerland\label{inst:sednai}
        \and
        Ru{\dj}er Bo\v{s}kovi\'{c} Institute, Bijeni\v{c}ka cesta 54, 10000 Zagreb, Croatia\label{inst:zagreb}\vfill
    }

   \date{Received date; Accepted date}

 
  \abstract
   {As part of \Gaia\ Data Release 3 (\GDR3), epoch photometry has been released for 1.2 million sources centred on M31. This is a taster for \Gaia\ Data Release 4 where all the epoch photometry will be released.}
   {In this paper the content of the \Gaia\ Andromeda Photometric Survey is described, including statistics to assess the quality of the data. Known issues with the photometry are also outlined.
   }
   {Methods are given to improve interpretation of the photometry, in particular, a method for error renormalization. Also, use of correlations between the three photometric passbands allows clearer identification of variables that is not affected by false detections caused by systematic effects.}
   {The \Gaia\ Andromeda Photometric Survey presents a unique opportunity to look at \Gaia\ epoch photometry that has not been preselected due to variability. This allows investigations to be carried out that can be applied to the rest of the sky using the mean source results. Additionally scientific studies of variability can be carried out on M31 and the Milky Way in general.}
   {}

   \maketitle
%
\section*{Note to the editor and referees}

\textcolor{red}{This manuscript is one of the official papers planned to accompany \GDR{3} in a special issue of A\&A. As agreed with A\&A, and as for previous releases, minor changes to figures and tables may happen due to last-minute changes in the data release. At this stage, however, we do not expect expect any change at all. 
}

\section{Introduction}\label{Sect:Introduction}

\Gaia\ is one of the most ambitious, diverse and demanding projects of the ESA Astrophysics Science programme that is in operation. From early on, the data processing and analysis of the \Gaia\ data have been recognised to be a challenge of the highest order. More than 450 people have gathered to achieve this enormous task. Ten years after the launch, investigations of alternative algorithms and software development are still ongoing to achieve significant improvements of the data products.
In order to mitigate the risk and satisfy the scientific community, the approach has been to release \Gaia\ data in an iterative manner: at each release more data has been processed, there are a larger number of sources with more diverse data products. The data behaviour is better understood and more and more effects are taken into account and therefore the calibrations are also improved. The feedback from the scientific community is also important in this process. The \Gaia\ Andromeda Photometric Survey (GAPS) is such an early release for the epoch photometry.

In the data release plans, the intention for \GDR3 was to only release data for the entire catalogue that had been averaged over many observations. The equivalent epoch data, sometimes referred to as
time series, would only be published in the fourth data release,  not expected before the end of 2025. However, the iterative approach could also be taken for the time domain measurements of \Gaia. In the first data release, \G-band epoch photometry was released for 3194 variable stars. In the second data release, this number increased to half a million variable stars. Now with the third data release about 10 million variables will be published with their time series. In this approach, it was thought that releasing epoch photometry for all sources, variable and constant, from a limited region of the sky would help the community understand the strengths and limitations of the \Gaia\ epoch photometry.

The paper outline is as follows: 
 Sect.~\ref{Sect:FieldChoice} describes the choice of the field for this survey; Sect.~\ref{Sect:Data} describes the data; in Sect.~\ref{Sect:Stats} the overall statistics is described; Sect.~\ref{Sect:KnownIssues} goes through some of the issues that remain in the data and Sect.~\ref{Sect:Examples} gives a few simple examples of what can be done with the data.
As with many large missions, \Gaia\ has many acronyms. Table~\ref{Tab:acronyms}  in the Appendix lists the ones used in this paper.

\section{Choice of field}\label{Sect:FieldChoice}
Several fields were studied to determine if they would be suitable for a data release. Among them were the Andromeda galaxy (M31), the Ursa Minor dwarf Spheroidal (UMi dSph), the open cluster NGC 2516 (a well-studied intermediate-age cluster), the Kepler field, and the two PLATO Long-duration Observation Phase fields \citep{PlatoFieldSelection}. 

The \Gaia\ scanning law is peculiar and these fields are sampled very differently. For example, the locations of Ursa Minor dwarf Spheroidal and NGC 2516 are close to the ecliptic poles and are benefiting from the very specific Ecliptic Pole Scanning Law (EPSL) from the first month of the operations before the spacecraft started its Nominal Scanning Law (NSL) \citep{DR1Mission}. Fields observed with the EPSL were observed very frequently during this period and 
are therefore excluded, as they would not be representative of the typical \Gaia\ sampling. The PLATO and Kepler fields are in a region with low number of measurements. The Andromeda galaxy instead is in a region where the number of scans varies significantly within the range covered by \Gaia\ due to its scanning law. The 10th and 90th percentiles of this distribution are 10 and 57 observations. Furthermore M31 encompasses regions of different densities, including crowded areas, so the community will be able to evaluate some spurious variability effects due to the crowding. Finally, with a radius of 5.5 degrees centred on M31 (RA 10.68333$^\circ$, Dec 41.26917$^\circ$), this field combines in addition to stars from the Andromeda galaxy, a large number of Milky Way stars that result in a HR diagram where all sequences are well populated. 

\section{Data description}\label{Sect:Data}

The data set contains epoch photometry in \G, \BP\ and \RP\ for 1\,257\,319 sources. The \G\ photometry is the Field of View (FoV) average, so there is only one \G\ value per transit. This data is contained within a DataLink Massive data base associated with the \GDR{3} archive. This can be accessed from the archive query results by clicking on the DataLink symbol (two chain links) and selecting the appropriate data from the pop-up window. 

There are three data structures that can be selected. 
The RAW data structure option will result in one file with one row per source with arrays for each field. Element $i$ of each array contains data for the $i$'th transit. This data structure is described in the online DR3 documentation\footnote{See \webref{https://archives.esac.esa.int/gaia}{https://archives.esac.esa.int/gaia}}
in Sect.~20.7.1. The other two data structures, INDIVIDUAL and COMBINED, have one row per transit and passband type. INDIVIDUAL has one file per source, while COMBINED has all the data within one file. Table~\ref{Tab:DataDescription} gives details about the fields of the epoch photometry contained within the INDIVIDUAL and COMBINED data structures. Note that \verb+rejected_by_photometry+ and \verb+rejected_by_variability+ are independent of each other and are the result of different processes.

To identify which sources are part of GAPS, the main \texttt{gaia\_source} archive table can be queried by checking
\verb+in_andromeda_survey = 't'+. Other source data can also be extracted at this time. Note that there is a limit associated with the archive DataLink service and only 5\,000 sources can be extracted in a single query. An example is given in App.~\ref{App:Downloading} of how to automate this and extract more than this limit.

\begin{table*}
 \caption{General description of the epoch photometry fields  of the \GDR{3} archive (for the INDIVIDUAL and COMBINED data structures).}
 \label{Tab:DataDescription}
 \centering
 \begin{tabular}{l l l}
  \hline\hline
  Name & Description & Units/Further Notes \\    
  \hline
  \verb+source_id+ & \Gaia\ DR3 Source ID & See Sect.~20.1.1 of documentation for encoding\\
  \verb+transit_id+ & Transit ID & See Sect.~20.7.1 of documentation for encoding\\
  \verb+band+ & Observation passband & G, BP or RP\\
  \verb+time+ & Time of observation & Barycentric JD in TCB - 2455197.5 (day)\\
  \verb+mag+ & Magnitude of epoch & Vega scale\\
  \verb+flux+ & Flux of epoch & e$^-$/s\\
  \verb+flux_error+ & Error on flux & e$^-$/s \\
  \verb+flux_over_error+ & Signal to noise ratio & \\
  \verb+rejected_by_photometry+ & Rejected by photometric processing (CU5) & True/False (unavailable, rejected or negative flux) \\
  \verb+rejected_by_variability+ & Rejected by variability processing (CU7) & True/False (rejected)\\
  \verb+other_flags+ & Additional processing flags & Coding described in App.~\ref{App:OtherFlags} \\
  \verb+solution_id+ & Solution ID & \\
  \hline
 \end{tabular}
\end{table*}

As reported in \cite{EDR3Phot}, when generating mean photometry in the processing leading to \EDR3, calibrated epoch fluxes with values lower than 1 e$^-$/s were rejected. A similar threshold was set for epoch photometry entering the archive at 0 e$^-$/s, i.e. only positive values were considered valid.
Ideally, the negative fluxes should have been retained since they are equally valid. For sources with very low flux, the error distribution is to all practical purposes Gaussian, however, when the fluxes are transformed to magnitudes, they lose this property. Additionally, negative fluxes, if they had been retained, would have undefined magnitudes. 
An alternative way, not used here, is to use an inverse hyperbolic sine function instead of a logarithmic transform, as proposed by \cite{LutponEtal1999}. Such transformation allows negative values of flux. 
Care must therefore be used when using the magnitudes, since there will be transits with very large magnitudes, especially for \BP\ and \RP, corresponding to 
flux values close to zero. These are well beyond the nominal detection capabilities of \Gaia.

\begin{figure*}
  \resizebox{\hsize}{!}{\includegraphics{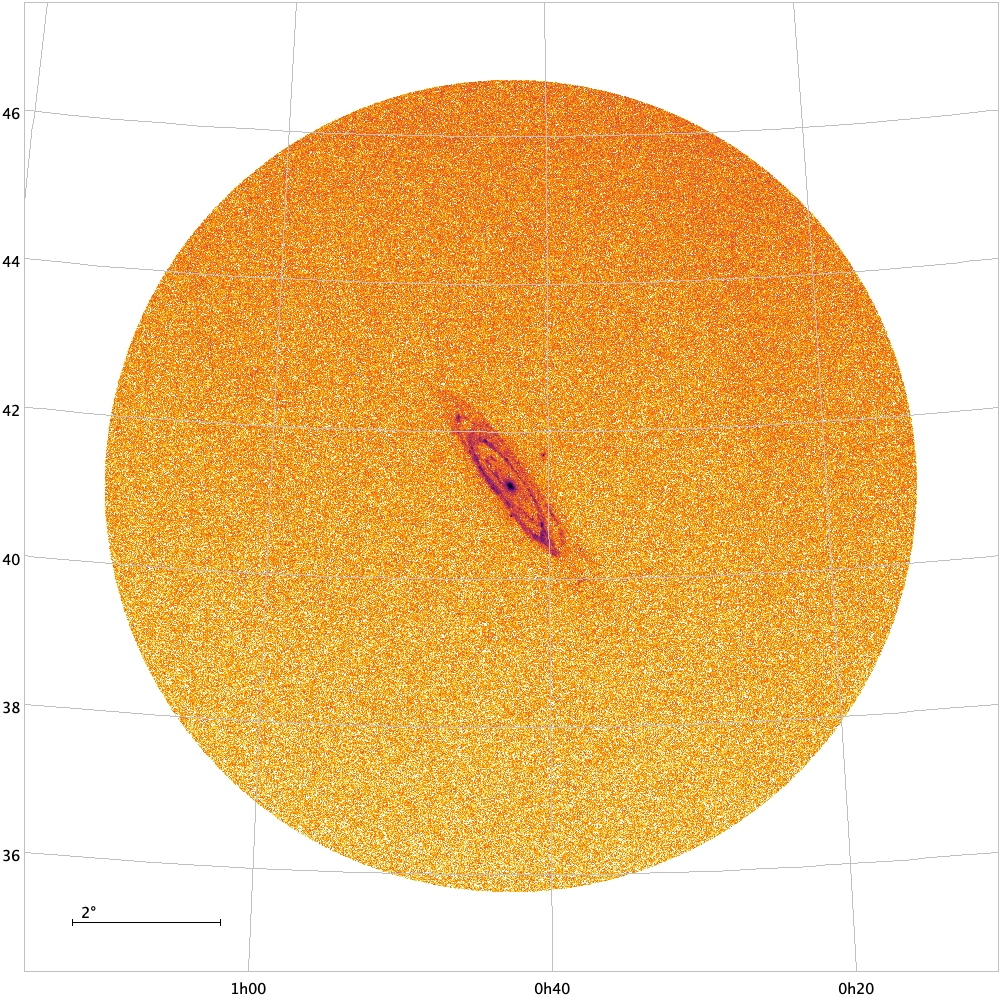}}
  \caption{The overall area covered by this survey in equatorial coordinates. As with many sky plots from \Gaia, this is not an image but a diagram showing sources identified by \Gaia. The density scale is logarithmic. Also visible are M32 and M110.}
  \label{Fig:SkyAll}
\end{figure*}


The sky distribution of the sources in this survey is shown in Fig.~\ref{Fig:SkyAll}.

Three of the fields in Tab.~\ref{Tab:DataDescription} can be further decoded to generate information that might be of use to the user. 

The first quantity is the \Gaia\ transit identifier containing information on the FoV, CCD Row, On-Board Mission Time (OBMT) and the AC position of the window. Details on how to decode this can be found in Sect.~20.7.1 (\verb+EPOCH_PHOTOMETRY+) of the \GDR{3} documentation. 
The additional processing flags can also be decoded using the description given in the Section 19.6.1 of the documentation. One of the bits of this flag indicates if \G\ band flux scatter is larger than expected by the photometry processing. If this is set for a significant number of the \G\ epochs of a source, this could indicate very short timescale variability. However, for some magnitude ranges, this could be due to uncalibrated systematic effects, such as magnitude terms, see Sect.~\ref{Sect:MagTerms}. Most of the bits in this flag mainly describe which CCD data was rejected or unavailable in forming the \G\ FoV average. 

Finally, the source identifier is described in Section 20.1.1 (\verb+GAIA_SOURCE+) of the \GDR{3} documentation. Contained within this number is a level 12 HEALPix index number which gives approximately the position of the source to the nearest arcmin.

\section{General statistics}\label{Sect:Stats}

\begin{figure}
  \resizebox{\hsize}{!}{\includegraphics{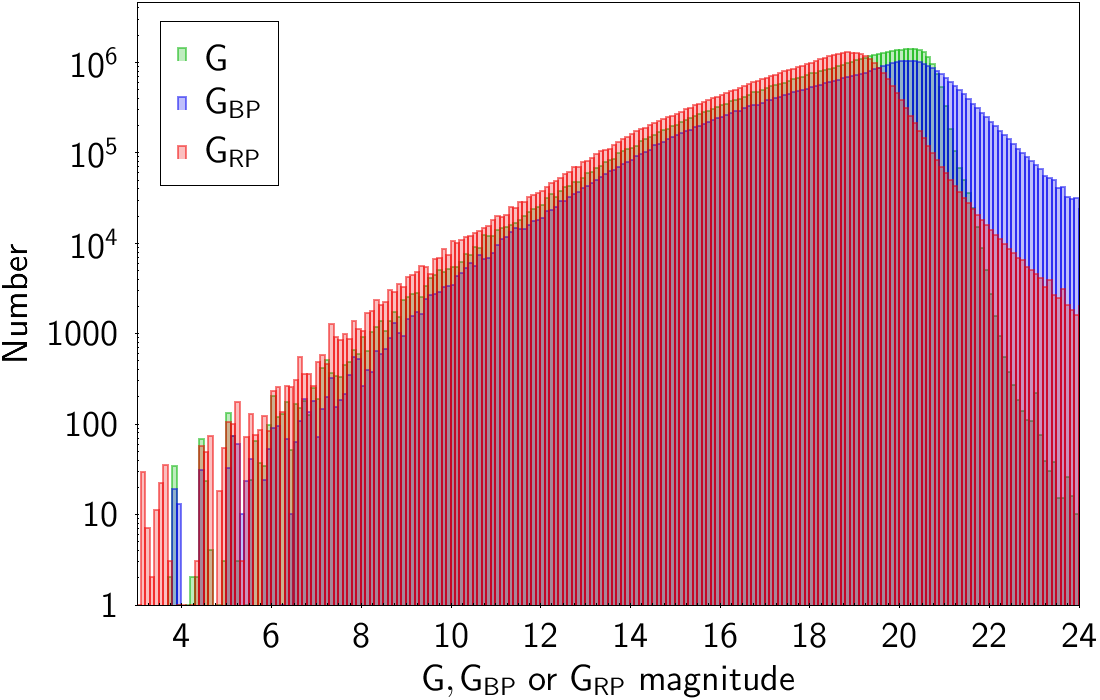}}
  \caption{Number of observations as a function of \G, \BP\ and \RP\ magnitude.}
  \label{Fig:StarCountsEpoch}
\end{figure}
 
Figure~\ref{Fig:StarCountsEpoch} shows the number of observations at each magnitude for \G, \BP\ and \RP. The peaks of these are approximately at 20.2, 20.2 and 18.9 mag respectively. The broadness of the peaks, extending well into the faint end, are an effect of the large uncertainties for faint sources, especially for \BP\ and \RP. Additionally, the asymmetric and non-linear nature of the transformation from flux to magnitude increases the number of extremely faint epochs. No signal-to-noise filter has been applied to this survey to avoid biasing any investigations that the users might want to carry out. 

\begin{figure}
  \resizebox{\hsize}{!}{\includegraphics{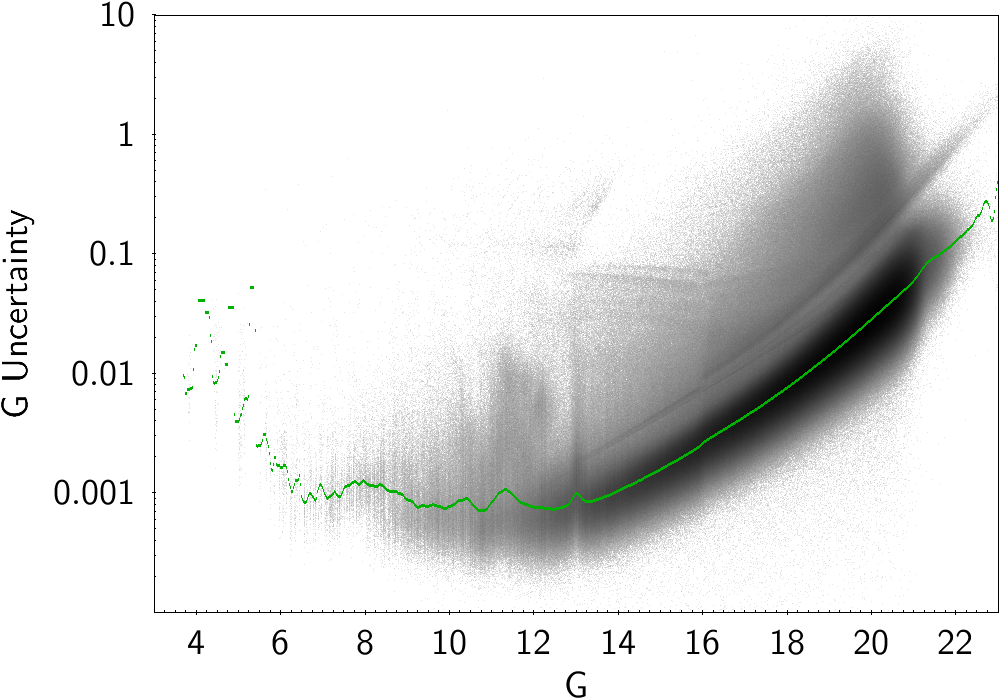}}
  \caption{Distribution of the uncertainty on \G\ in magnitudes as a function of magnitude. The line shows the median of the distribution. The density colour scale is logarithmic.}
  \label{Fig:ErrG}
\end{figure}
 
\begin{figure}
  \resizebox{\hsize}{!}{\includegraphics{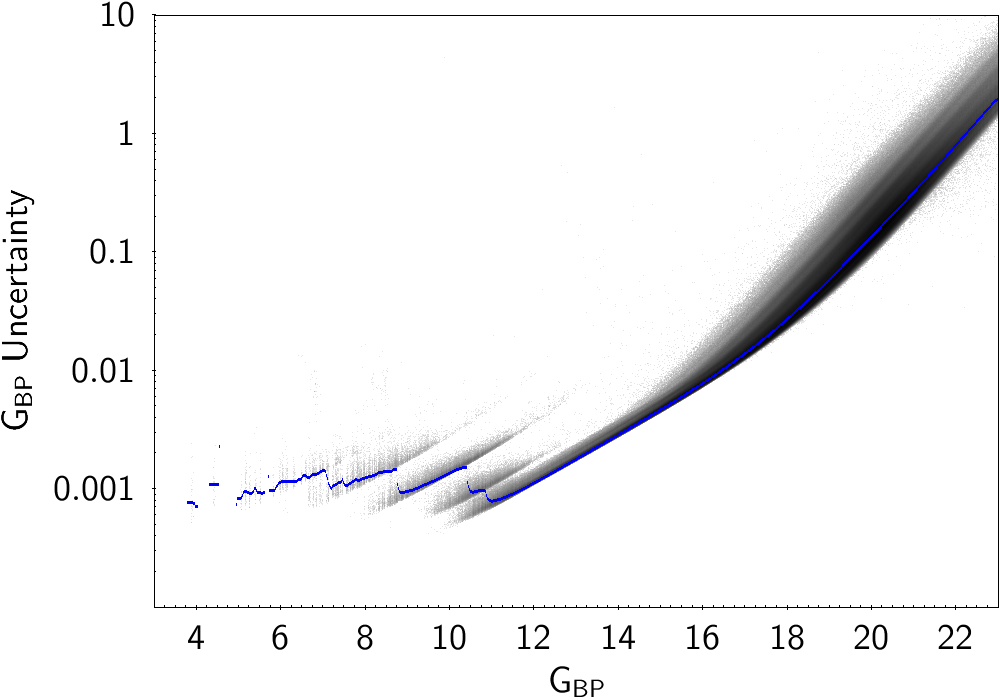}}
  \caption{As Fig. \ref{Fig:ErrG}, but for \BP.}
  \label{Fig:ErrBP}
\end{figure}
 
\begin{figure}
  \resizebox{\hsize}{!}{\includegraphics{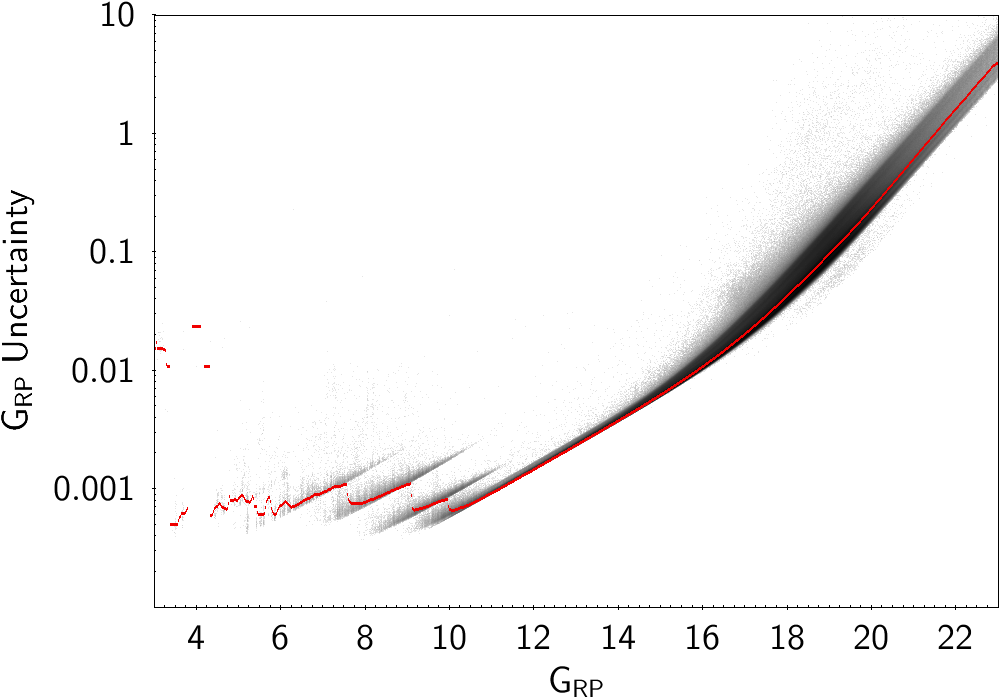}}
  \caption{As Fig. \ref{Fig:ErrG}, but for \RP.}
  \label{Fig:ErrRP}
\end{figure}
 
\begin{figure}
  \resizebox{\hsize}{!}{\includegraphics{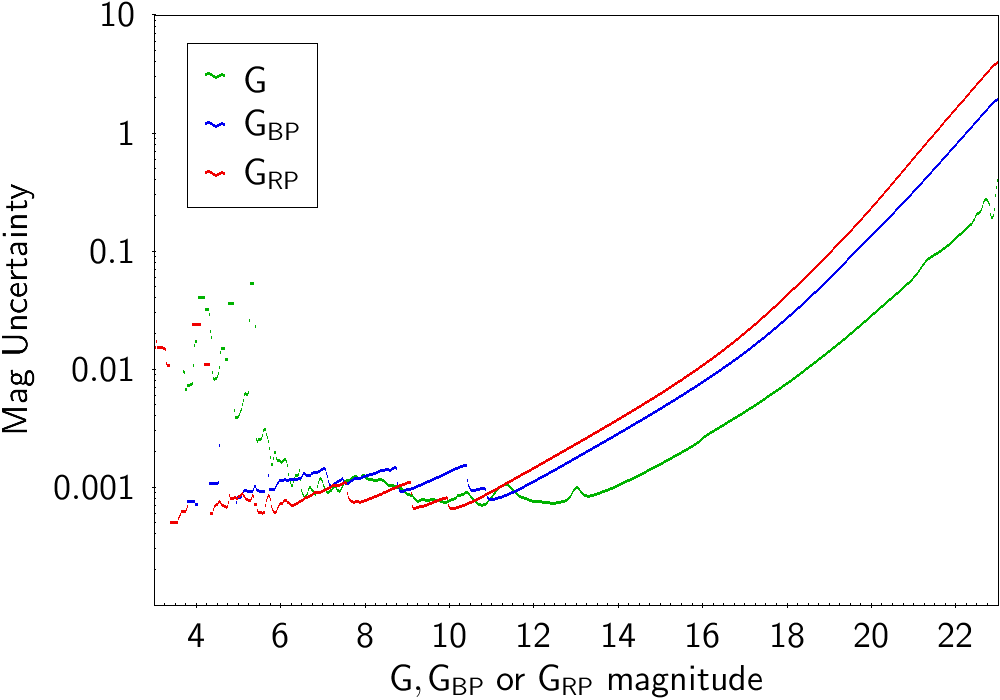}}
  \caption{The median of the \G, \BP\ and \RP\ uncertainties  in magnitudes as a function of magnitude.}
  \label{Fig:ErrAll}
\end{figure}
 
The uncertainties are shown in Figs. \ref{Fig:ErrG} to \ref{Fig:ErrRP}. The \G\ fluxes are formed from a weighted mean of  up to 9 AF CCDs forming the transit. The uncertainties reflect this. The formulae for calculating the weighted mean and uncertainty can be found in \cite{DR1PhotPrinciplesErratum}. This means that some features, such as gating, that are seen in the equivalent \BP\ and \RP\ plots, are smoothed out. The ridge observed at the faint end, about a factor 5 above the median line, is formed by transits that only contain one CCD measurement (AF1). This can be identified from the flags described in App.~\ref{App:OtherFlags}, indicating which AF CCDs contributed to the mean \G\ flux value.

Some of the features seen above the median line in the \G\  magnitude range 13 to 16 are caused by variation of the photometry within the transit. This could be due to variability or uncalibrated systematic effects affecting one or more of the 
CCD measurements within the transit.

At the bright end, \G$<$12, the bumps seen are caused both by the different effective exposure times caused by gating and by saturation features. Although gating should mitigate most of the effects of saturation, the on-board choice of gate is affected by on-board photometric errors and is therefore not always optimal thus causing some saturation to occur. This is different from CCD to CCD and therefore additional scatter is observed.

The \BP\ and \RP\ uncertainties have an easier structure to explain. Brighter than about magnitude 11, the features are all caused by gating. The effect of this is for the transits to have different effective exposure times and therefore uncertainties. Note that the selection of the gate to be used is done by an on-board magnitude estimate which is quite noisy (brighter than about \G=12, this is about 0.5 mag). This explains why there is quite an overlap in magnitude between the different gated observations.

Brighter than about magnitude 16, the gradients in these plots are all very similar and indicate that the uncertainty is source-limited \ie{ limited by the signal-to-noise of the source}. Fainter than this, the gradient changes showing the change into a sky-limited regime.

Figure \ref{Fig:ErrAll} shows the median of the distribution for all three passbands on the same plot to ease comparison.

 \begin{figure}
  \resizebox{\hsize}{!}{\includegraphics{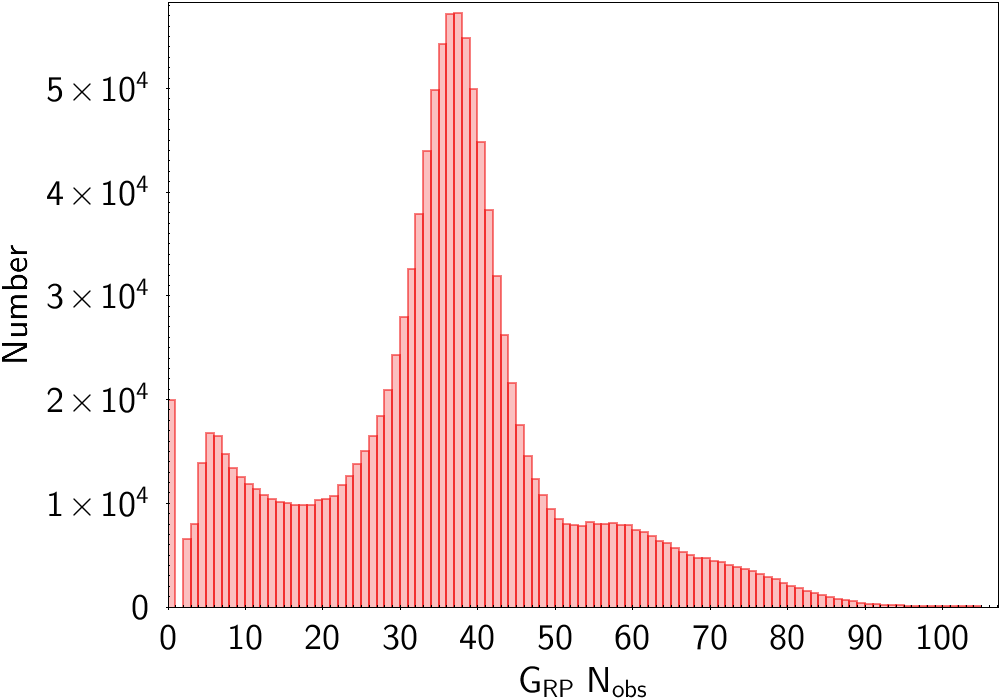}}
  \caption{The total number of \RP\ observations for each source within the survey.}
  \label{Fig:HistRPNobs}
\end{figure}
 
 \begin{figure}
  \resizebox{\hsize}{!}{\includegraphics{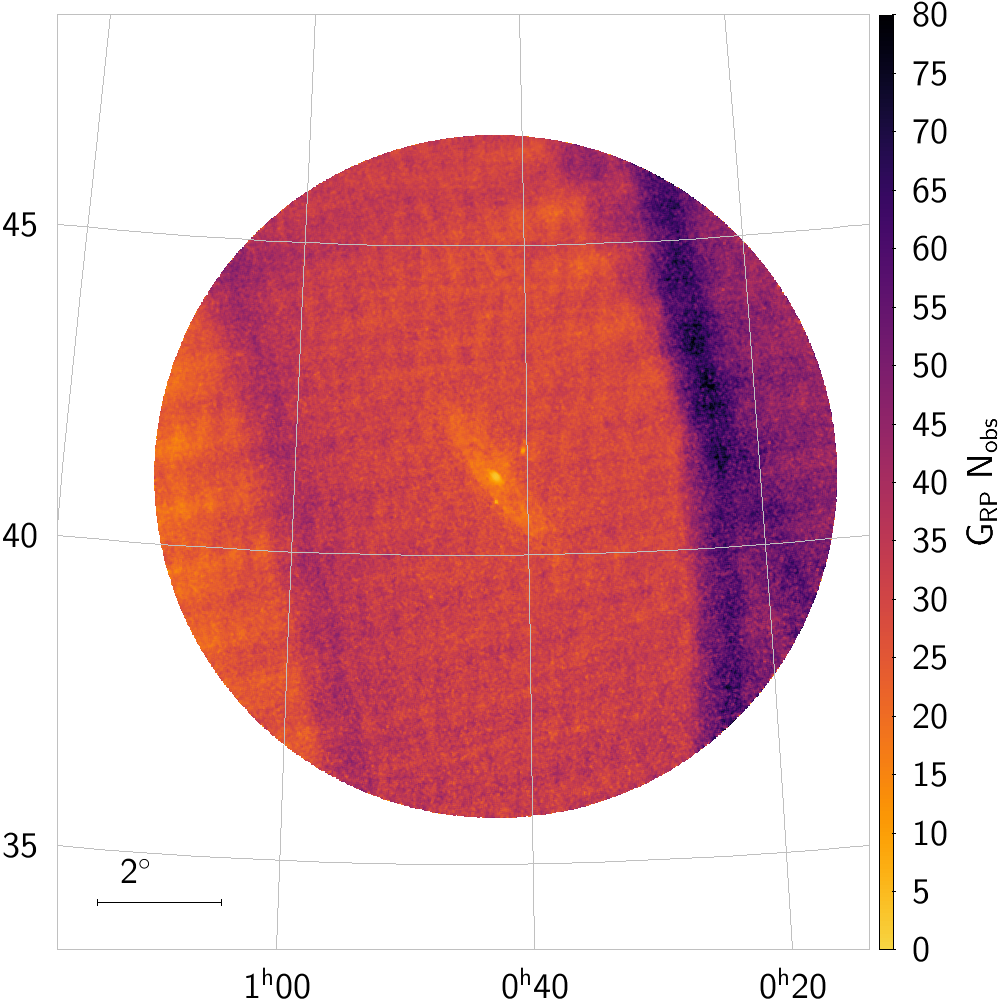}}
  \caption{The sky distribution of the sources  in equatorial coordinates weighted by the number of \RP\ observations.}
  \label{Fig:SkyRPNobs}
\end{figure}
 
The total number of observations per source is shown in Fig. \ref{Fig:HistRPNobs} and their sky distribution in Fig. \ref{Fig:SkyRPNobs}. The \RP\ values are shown in these plots as representative of the number of transits in each passband.

The majority of sources have between 30 and 45 observations, but a reasonable number, 15\%, have more than 50 observations. These are located in the stripes seen in the sky distribution and are due to the scanning law of \Gaia. The central region of M31 has very few observations in comparison, due to crowding causing the observations to fail for a number of reasons. For the \G\ observations, the Line Spread Function (LSF, \citealt{EDR3PSF}) fits fail due to the presence of multiple sources in the window. \BP\ and \RP\ observations have much larger windows, which often overlap in crowded regions and are therefore truncated on-board. These have not been processed for \GDR{3}. The occurrence of overlap and therefore truncation depends not only on stellar density but also the scanning direction with respect to the location of the sources in the sky. For this reason, most sources will still get some useful observations but the average number of epochs per source is significantly reduced in such areas.

 \begin{figure}
  \resizebox{\hsize}{!}{\includegraphics{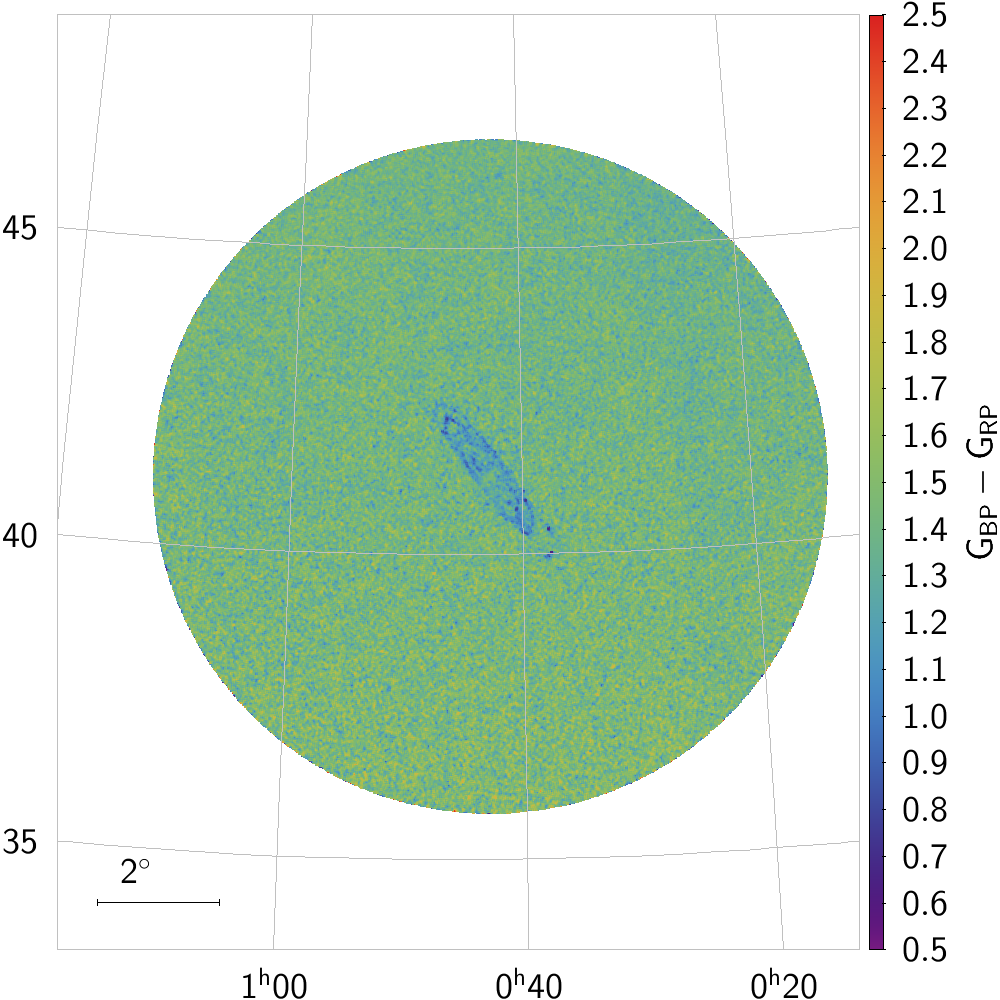}}
  \caption{The sky distribution of the sources  in equatorial coordinates weighted by the colour of the source.}
  \label{Fig:SkyColour}
\end{figure}

Figure \ref{Fig:SkyColour} shows the colour distribution in GAPS field. The average colour away from M31 is 1.4 in \BP-\RP, whereas in the spiral arms of M31 the average colour is 1.0. This reflects the brighter population of M31.

\begin{figure*}
  \resizebox{\hsize}{!}{\includegraphics{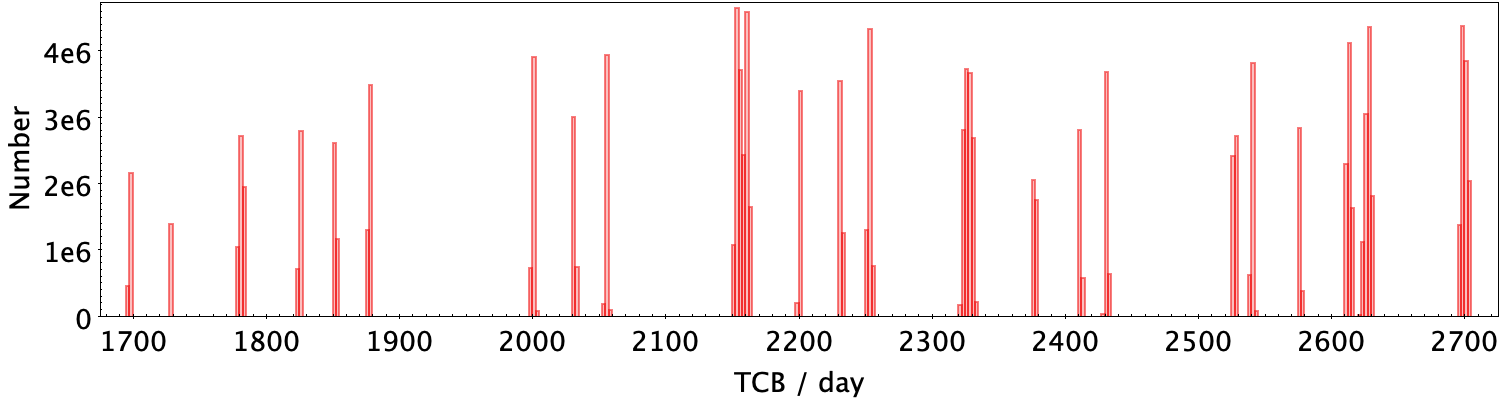}}
  \caption{Distribution in time (TCB days since 2010) of the epochs within the survey. The range of 1700 to 2700 TCB days since 2010 corresponds approximately to 1200 to 5200 OBMT rev, which is often used in other \Gaia\ papers.}
  \label{Fig:TimeDistribution}
\end{figure*}

Figure \ref{Fig:TimeDistribution} shows the time distribution of the epochs
as Barycentric Julian Day in Barycentric Coordinate Time (TCB) - 2455197.5 in days. This corresponds approximately to the range 2014-08-27 to 2017-05-24. As can be seen the distribution is highly irregular due to the \Gaia\ scanning law.

\section{Known issues with the published photometry}\label{Sect:KnownIssues}
\subsection{Error renormalization}\label{Sect:ErrorRenormalization}

It is often the case that the estimation of errors is as difficult to get right as the main data. 
A correct estimation of the errors on the single transits is very important, since modelling the data often relies on errors for weighting the data. Usually errors are underestimated in comparison to the observed scatter due to uncalibrated systematic errors. Distinguishing between some systematic errors and random ones may not be important in many modelling cases where the model doesn't use the parameter driving the systematic. For example, if there were systematic differences between CCD Rows or FoV, these would not be relevant for scientific modelling such as light curve fitting apart from an apparent increase in the size of the errors.

\begin{figure*}
  \resizebox{\hsize}{!}{\includegraphics{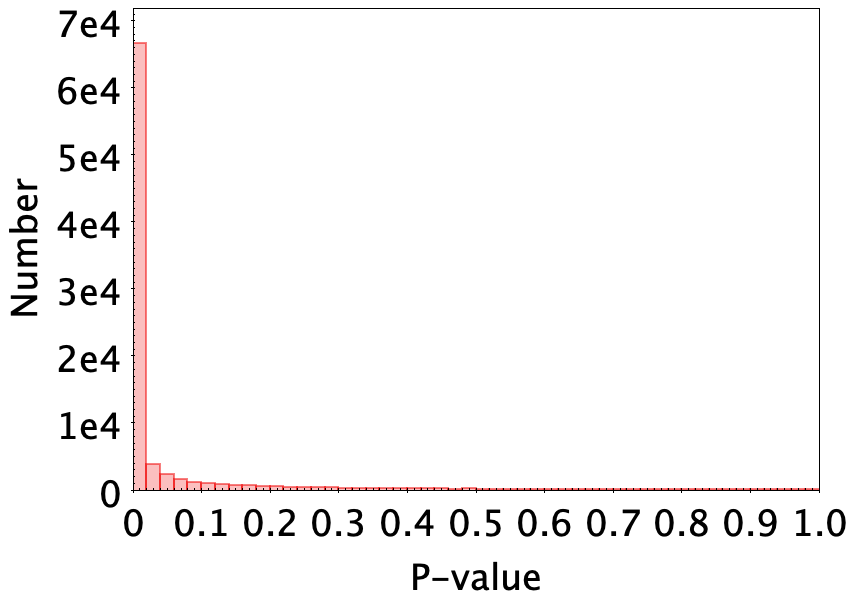}\includegraphics{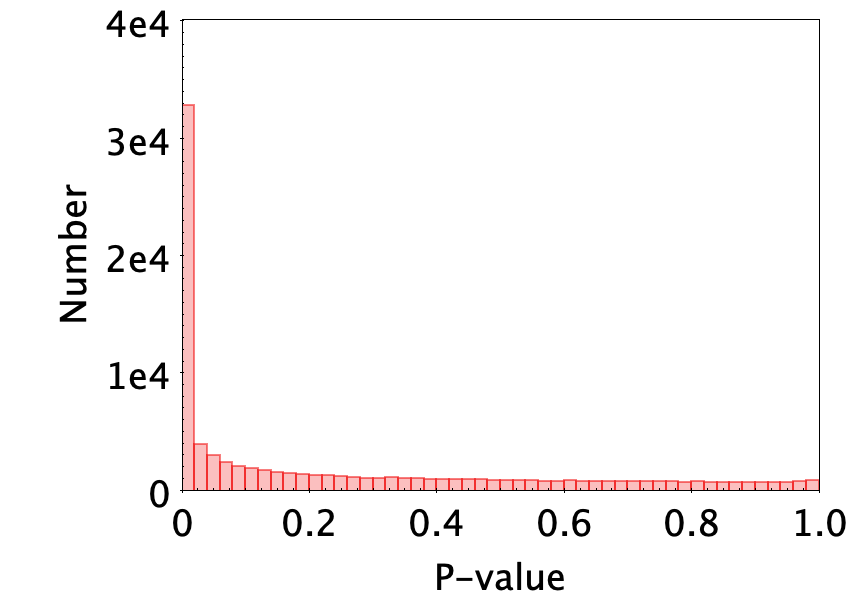}\includegraphics{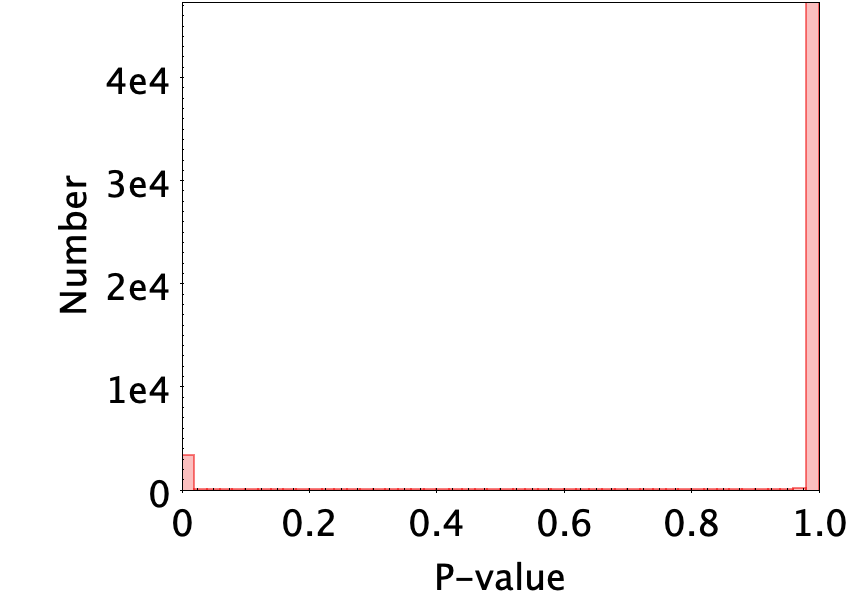}}
  \caption{\G\ band P-value distributions for sources in the range 16$<$\G$<$17 for three cases in which the following magnitude errors were added in quadrature to the errors provided in the archive: 0.0 (left panel), 0.0019 (central panel), 0.01975 (right panel). The middle value corresponds to the selected additional error for this magnitude range.}
  \label{Fig:PValueDistributions}
\end{figure*}

In many cases, a simple investigation of the unit-weight residuals can give an indication as to the quality of the error estimates. This usually involves a modelling assumption, \eg the source is constant. A particularly useful technique is the use of P-values which effectively transforms the residuals into a flat distribution which can be interpreted more easily, see Eq.2 of \cite{DR1PhotValidation} for the conversion from $\chi^2$ to P-value. 
Indeed, in classical hypothesis testing \citep{1979ats..book.....K}, the P-value is used to accept or reject the null hypothesis. It is defined (for a unilateral test) as the probability of the random variable $X$ to be larger than the obtained value $x_o$.
$$
\mbox{P-value} = \int_{X_0}^{\infty}   X dX
$$
The random variable $X$ is assumed to follow the statistics of the null hypothesis.
In the case where all sources follow the null hypothesis, the distribution of P-value is flat.
As with a $\chi^2$ test, the P-value test is very sensitive since involves a quadratic residual. If a small fraction of the sources are variable and the uncertainties are well estimated, then the distribution of the P-values is flat for a large fraction of the P-values, and a peak is present near zero for the variable sources. However, it is not straightforward to disentangle the effect of variability from that of an unrepresentative error estimation. The left panel in Fig.~\ref{Fig:PValueDistributions} shows this issue (variability and error problems) where the original P-value distribution is shown for sources in this survey in the magnitude range 16$<$\G$<$17.

The usual approach for error renormalization is to scale the errors in some way. This is what was carried out for the astrometry in \GDR{2} \citep{DR1Astrometry} where this was the correct mitigation. This approach was attempted for the photometry, but no consistent results could be achieved. It is likely that the main problem affecting the unit-weight residuals of the photometry is caused by uncalibrated systematic effects. Since there are likely to be more than one of these, the combined effect is to effectively introduce an additional random Gaussian error, in agreement with the central limit theorem \citep{1977ats..book.....K}. This suggests that adding an error in quadrature to the formal error would be a better solution than scaling the errors. Since the size of the systematic errors for \Gaia\ is  likely to be a function of magnitude, the additive correction should be a function of magnitude. Corrections to the photometric errors have been computed independently for the three different passbands. An example of these corrections is shown in the last 2 panels of Fig.~\ref{Fig:PValueDistributions} where two different corrections have been added in quadrature to the formal errors before calculating the P-values. In the right-most panel, the additional error is too large causing an excess of high P-values. The addition of 0.0019 mag in the middle panel gives a reasonable solution where most of the distribution is flat and the peak at low P-values would be caused by variability only.

The features seen in Fig. \ref{Fig:PValueDistributions} lead to a simple algorithm to determine a reasonable additional error that could be added in quadrature to the data. For each magnitude range ($\pm$0.5 mag) and starting with a very high additional error, the P-value distribution is generated and the ratio of sources in the P-value ranges [0.8,0.9] and [0.9,1.0] calculated. Initially, there will be far more sources in the last bin since the additional error is too large. This is gradually decreased for each magnitude range until the number of sources with P-value between 0.9 and 1.0 is smaller than the number of sources in the range [0.8,0.9]. When this condition is met, the corresponding correction is the one to be adopted. The step size of the decrease is 0.05 mmag for \G\ and 0.2 mmag for \BP\ and \RP.

\begin{figure*}

  \resizebox{\hsize}{!}{
  \includegraphics{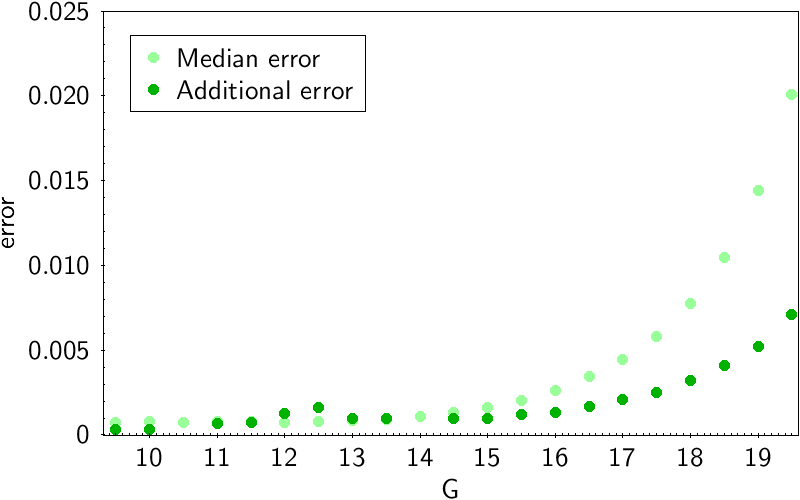}\qquad\includegraphics{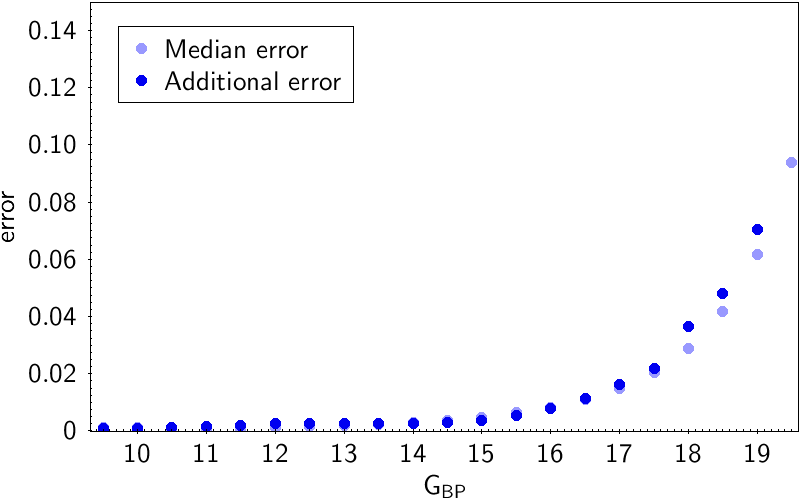}}
  \resizebox{\hsize}{!}{
  \includegraphics{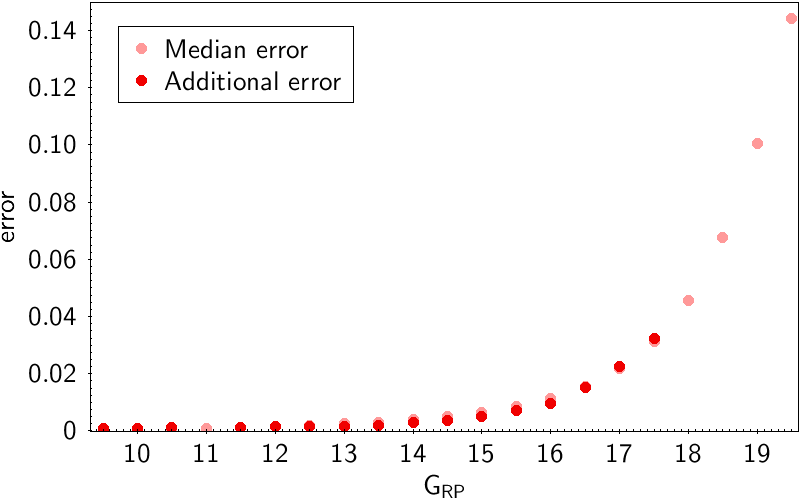}\qquad\includegraphics{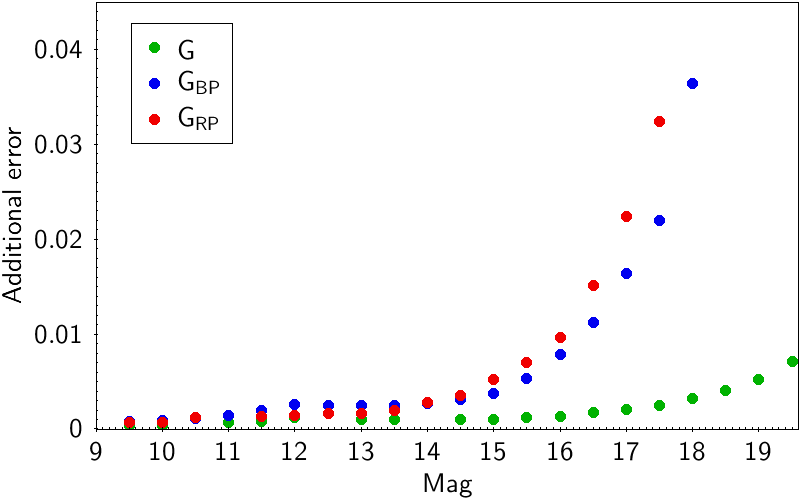}}
  \caption{The first three plots show the results for the error renormalization analysis for \G, \BP\ and \RP\ respectively. The errors are all in magnitudes. In these plots, the median quoted error is shown along with the additional error from this analysis. Note that the BP and RP plots have a larger ordinate axis maximum. The final plot shows the additional error for all 3 passbands at the same time to aid comparison. Missing points indicate where the algorithm has failed.}
  \label{Fig:PValueRenormalization}
\end{figure*}

The results of this analysis are shown in Fig.~\ref{Fig:PValueRenormalization} in comparison to the median quoted errors. Table~\ref{Tab:PValueRenormalization} shows the results as a function of magnitude. These can be interpolated for generating the appropriate error to add in quadrature for each transit. In the cases where the algorithm has failed, \eg too few data points, no value is given. Interpolation over these points can be carried out. Extrapolation is not advised and the end values should be used in these cases.

From Fig.~\ref{Fig:PValueRenormalization} and Tab.~\ref{Tab:PValueRenormalization} it can be seen that the \G\ additional errors are much smaller than those of \BP\ and \RP. This is due to the G values being an average of up to 9 values. For the additional errors, the \G\ ones are smaller than the median error by around a factor 2, whereas the \BP\ and \RP\ ones are around the same size as the median quoted errors.

\begin{table}
 \caption{Additional single transit errors for \G, \BP\ and \RP\ as a function of magnitude. The dashes are where the algorithm has failed.}
 \label{Tab:PValueRenormalization}
 \centering
 \begin{tabular}{r c c c}
  \hline\hline
  Mag & \G & \BP & \RP \\    
  \hline
 9.5 & 0.0003 & 0.0008 & 0.0007 \\
10.0 & 0.0003 & 0.0009 & 0.0007 \\
10.5 &     -- & 0.0011 & 0.0012 \\
11.0 & 0.0007 & 0.0014 &     -- \\
11.5 & 0.0008 & 0.0020 & 0.0013 \\
12.0 & 0.0013 & 0.0026 & 0.0014 \\
12.5 & 0.0016 & 0.0025 & 0.0016 \\
13.0 & 0.0009 & 0.0025 & 0.0016 \\
13.5 & 0.0009 & 0.0025 & 0.0020 \\
14.0 &     -- & 0.0027 & 0.0028 \\
14.5 & 0.0010 & 0.0031 & 0.0035 \\
15.0 & 0.0010 & 0.0037 & 0.0052 \\
15.5 & 0.0012 & 0.0053 & 0.0070 \\
16.0 & 0.0014 & 0.0079 & 0.0096 \\
16.5 & 0.0017 & 0.0112 & 0.0151 \\
17.0 & 0.0021 & 0.0164 & 0.0224 \\
17.5 & 0.0025 & 0.0220 & 0.0324 \\
18.0 & 0.0032 & 0.0364 &     -- \\
18.5 & 0.0041 & 0.0480 &     -- \\
19.0 & 0.0052 & 0.0704 &     -- \\
19.5 & 0.0071 &     -- &     -- \\
20.0 & 0.0091 &     -- &     -- \\
  \hline
 \end{tabular}
\end{table}


\subsection{Magnitude-based systematics}\label{Sect:MagTerms}

Within the internal photometric calibrations, no terms depending on magnitude are used \citep{EDR3Phot}. This is because the reference photometry used for these calibrations is derived from the photometry itself in an iterative loop. Introducing a magnitude-dependent term into the calibration would cause convergence problems arising due to the overall system being degenerate.

Using the data in this survey, it is possible to see the scale of the magnitude dependent systematic effects in each of the three passbands by looking at the differential magnitude systematics. In  future processing cycles, these effects could be calibrated out once the mean reference photometry has been determined.

A number of effects can cause systematic deviations as a function of magnitude which can be very different between the \G\ passband data and that of \BP\ and \RP. For \G, the main effect comes from the fit of the LSF or PSF to the sampled data. If the calibration of the LSF/PSF is not perfect, then magnitude effects can arise due to the weighted nature of the fit.

The other significant effect comes from the calibration of the background. This affects \G\ in a similar manner to \BP\ and \RP. Problems with this calibration lead to a systematic effect at the faint end similar to a hockey stick.

For the \G\ photometry, an occasional systematic can be seen at around G=11 which is caused by saturation not mitigated by the gating strategy of \Gaia.

Note that these systematics will be different in each processing cycle since their cause is entwined with the different calibrations that have been carried out. With each processing cycle, the calibrations improve and the size of these magnitude terms are reduced.

\begin{figure}
  \resizebox{\hsize}{!}{\includegraphics{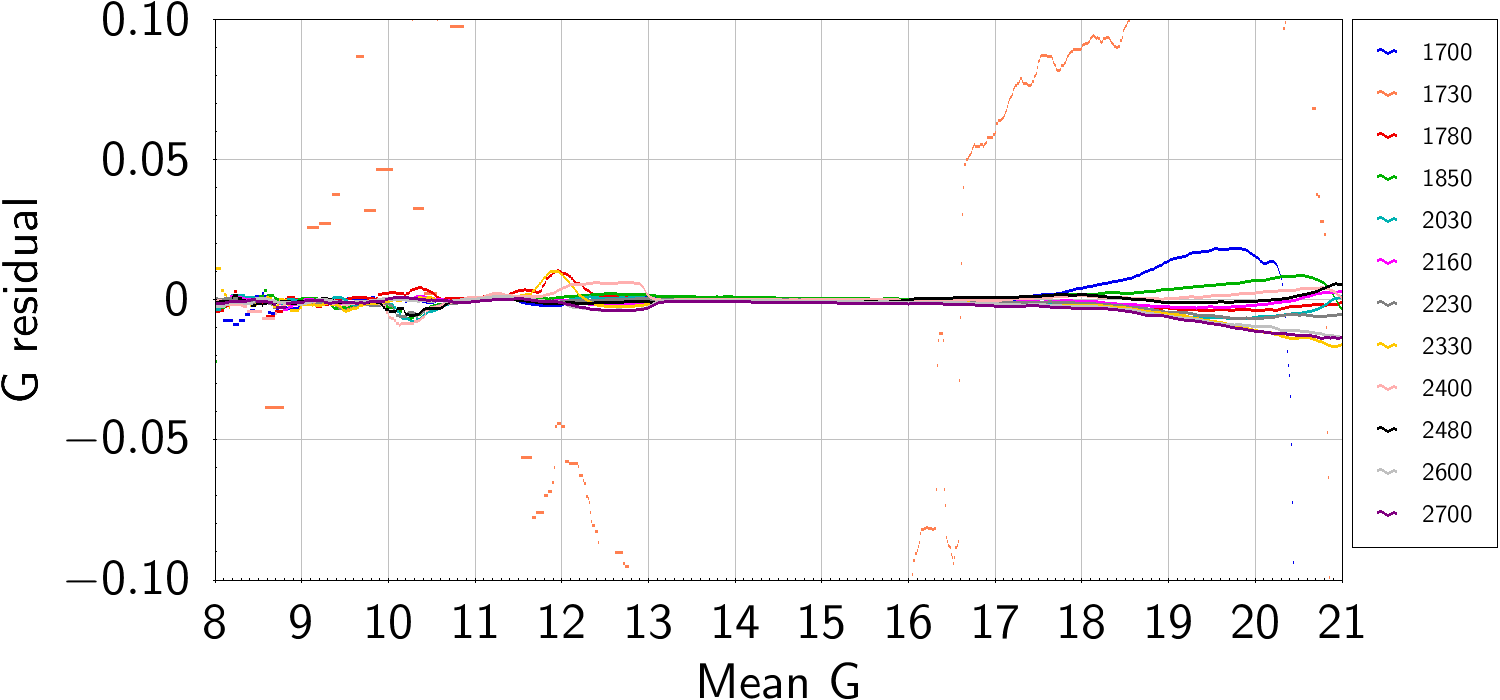}}
  \caption{Comparison between epoch and mean magnitudes for various time selections as a function of magnitude. The lines correspond to the medians of the distribution. The legend on the side indicates the time corresponding to the selection (TCB days since 2010). The data for this plot is from the Following FoV and Row 1.}
  \label{Fig:GmagTerms}
\end{figure}

\begin{figure}
  \resizebox{\hsize}{!}{\includegraphics{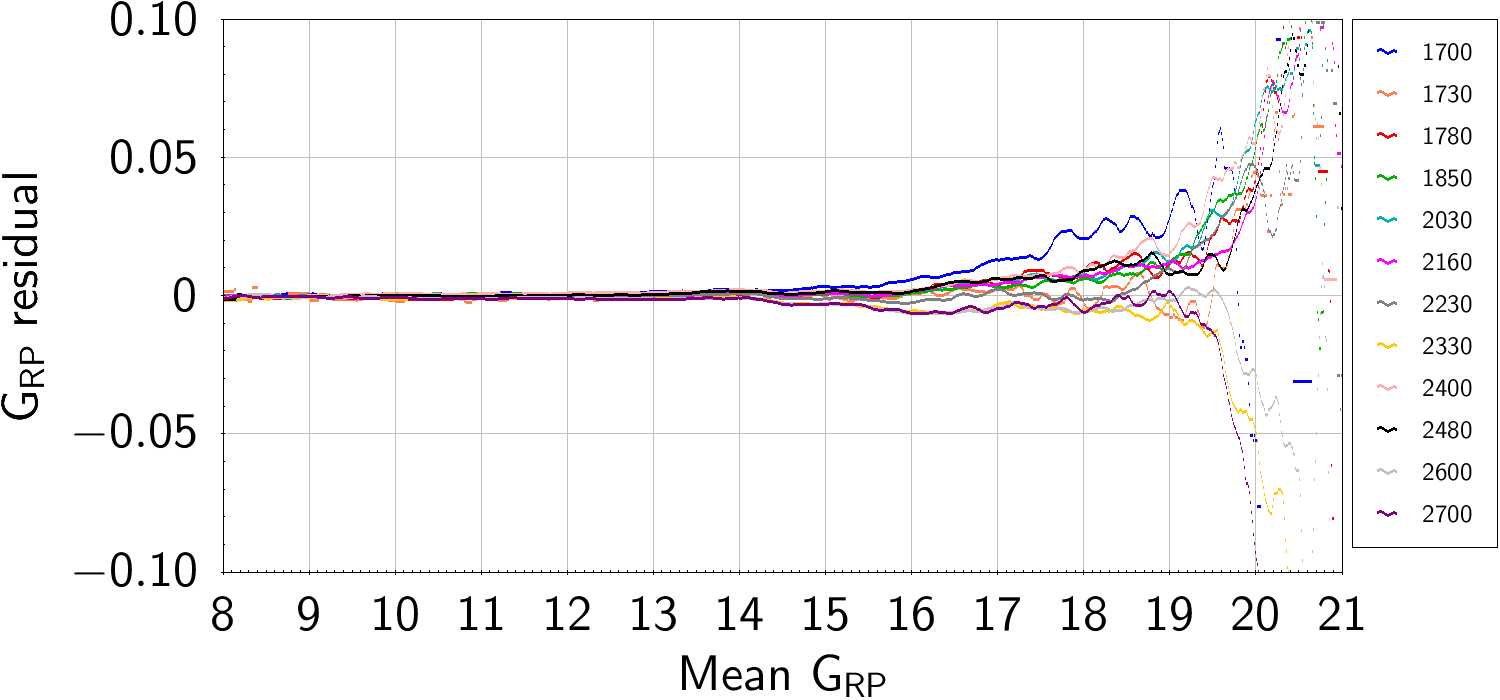}}
  \caption{As Fig.~\ref{Fig:GmagTerms}, but for \RP.}
  \label{Fig:RPmagTerms}
\end{figure}

Figure \ref{Fig:GmagTerms} shows the epoch \G\ residuals to the mean magnitude for the following FoV and Row 1 as a function of magnitude for different time selections. These correspond to the peaks seen in Fig.\ref{Fig:TimeDistribution}. Some of the narrower peaks have been grouped together to make the plot clearer. Only the medians of the distributions are shown.

The main two features that can be seen are the effects of saturation for $G<13$ and probable background subtraction issues at the faint end. These are clearly a function of time. No consistent pattern with time is evident.

Also seen in this plot is the outlier behaviour of the data around TCB=1730. The \G\ epochs for this period can deviate from the mean by a few 0.1 magnitude. This period was immediately following a decontamination event \citep{DR1Mission} and the image quality had not stabilized following the heating up of the focal plane. Thus the LSFs and PSFs generated for the time were not well suited for the data. Note that the \G\ data from this period is flagged as \verb+rejected_by_variability+. The corresponding \BP\ and \RP\ epochs are not similarly affected since their photometry is not determined by a profile fit, but effectively by aperture photometry, and is not as affected by image instability.

Figure \ref{Fig:RPmagTerms} shows the equivalent residuals for the \RP\ data. Here the only effect seen is at the faint end and is probably caused by difficulties with the background calibrations.
The \BP\ residuals show similar behaviour.

Comparisons between fields of view or CCD rows for the same time selection also show similar sized systematic trends. This indicates that these complex magnitude-based systematics depend on row, FoV and time. Note that the size of these systematic effects are a fraction of the size of the scatter, equivalent to the epoch uncertainty, usually less than 25\%.

\subsection{Crowding and background effects} 

\begin{figure}
 \resizebox{\hsize}{!}{\includegraphics{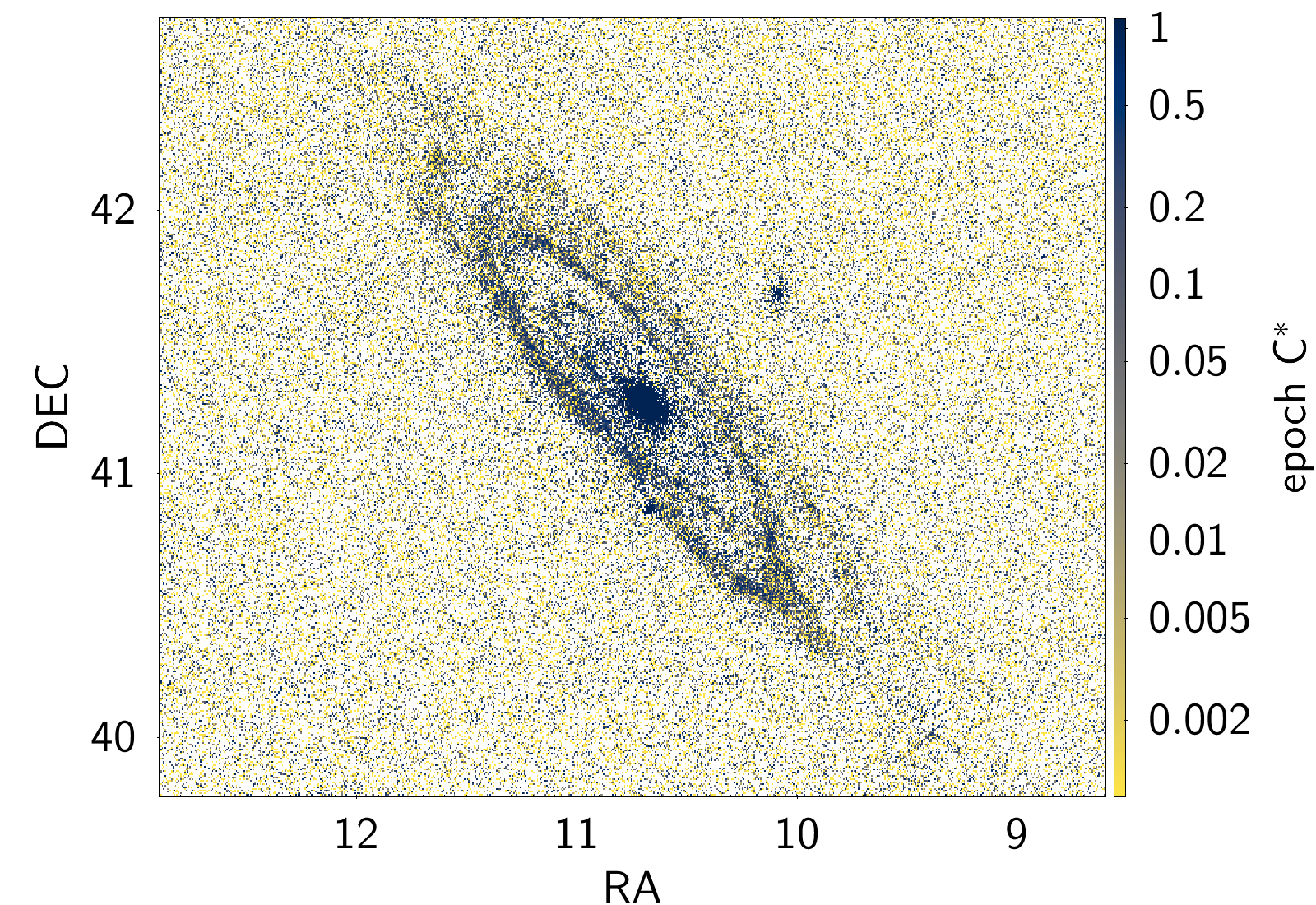}}
  \caption{Sky distribution of the corrected  \BP\ and \RP\ flux excess factor $C^*$, obtained from the epoch photometry.}
  \label{Fig:epochCstarSky}
\end{figure}

\begin{figure}
 \resizebox{\hsize}{!}{
  \includegraphics{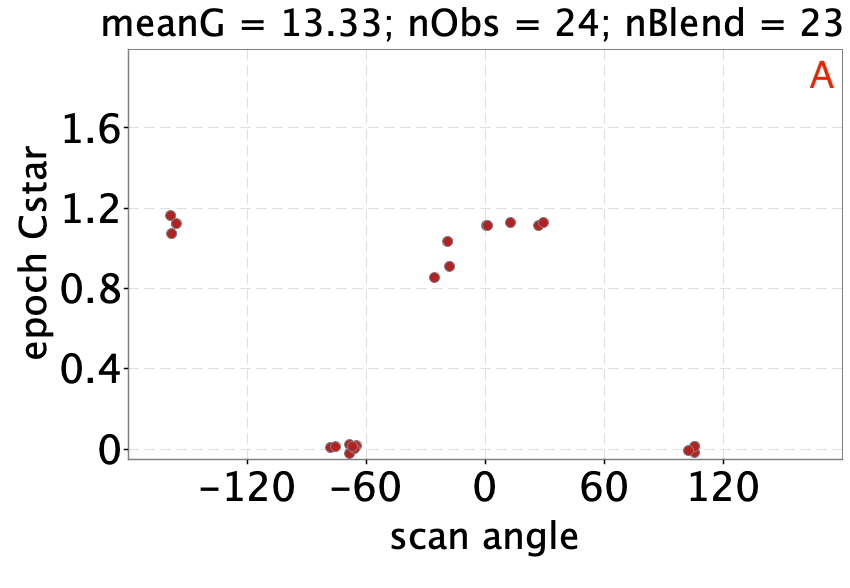}
  \includegraphics{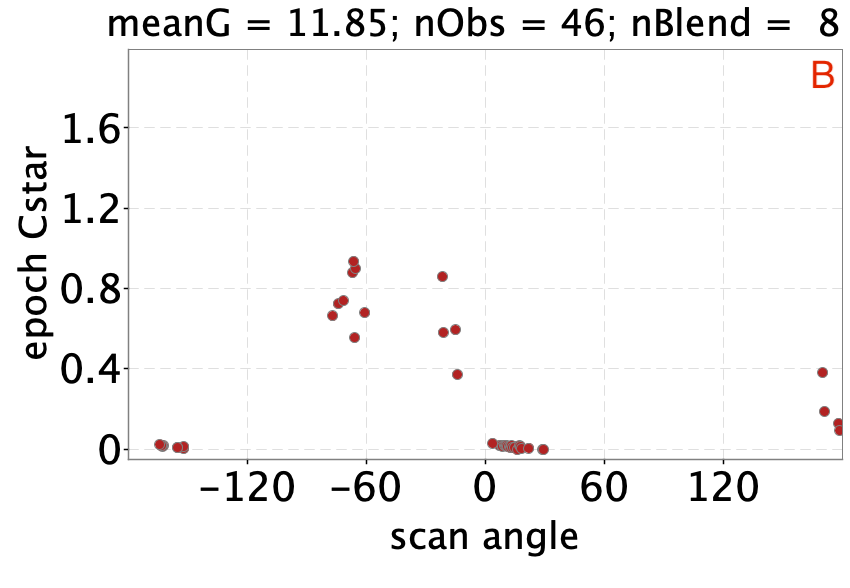}}
  \resizebox{\hsize}{!}{
  \includegraphics{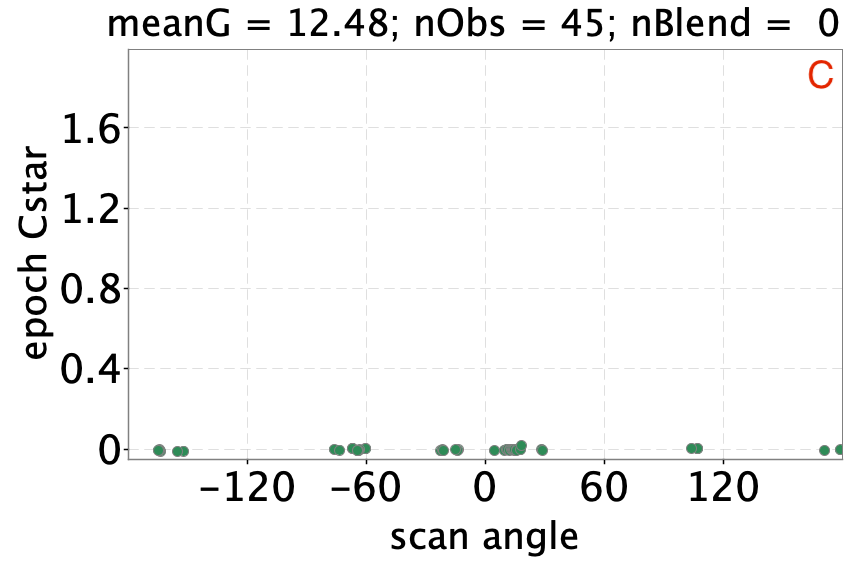}
  \includegraphics{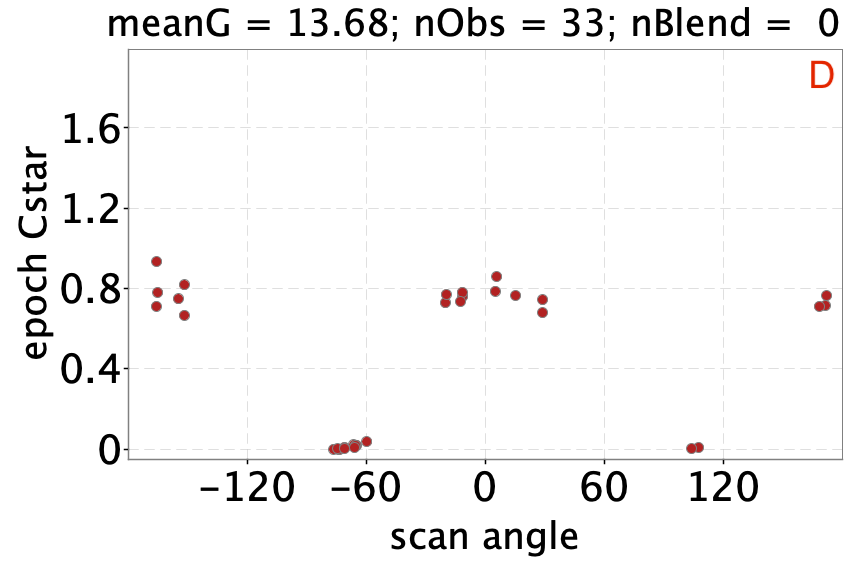}}
  \caption{Examples of \cxs variations with the scan angles (in degrees). In every panel the mean \G\ magnitude is indicated, as well as the total number of transits (nObs) and the number of blended transits (nBlend). See the text for a detailed explanation of the single cases.} 
  \label{Fig:epochCstarScanExamples}
\end{figure}

As described in \cite{EDR3Phot}, the corrected \BP\ and \RP\ flux excess factor was introduced as a consistency metric. The reader is reminded the definition of this quantity
\begin{equation}\label{eq:cxs}
\cxs=C-f(\bprp),
\end{equation}
where  $C=(I_{\textrm{BP}}+I_{\textrm{RP}})/I_G$ is the ratio between the sum of the BP and RP fluxes  and the G flux and $f(\bprp)$ is a function of the colour of the source.
Good and consistent photometry should have \cxs values around zero. In \GDR 3 the \cxs was calculated for all sources from their mean photometry. In the case of the GAPS dataset, it is also possible to calculate this from the epoch photometry to have an indication of its consistency. 
Figure \ref{Fig:epochCstarSky} shows the sky distribution, zoomed in on the Andromeda galaxy, of the epoch \cxs: in the centre of the galaxy and in the spiral arms  \cxs is clearly higher than the background, mainly due to crowding effects. At every epoch the scan angle changes, and depending on this, a source can be affected by neighbours stars in different ways, as the amount of contaminating flux varies.  
Figure \ref{Fig:epochCstarScanExamples} shows some examples of this effect: panel A is an example of a source that has almost all transits flagged as blended\footnote{The information about the number of blended transits comes from the main source catalogue (\texttt{phot\_bp\_n\_blended\_transits} and \texttt{phot\_rp\_n\_blended\_transits}).} but as the amount of contaminated flux varies with the scan angle, the crowding does not affect the photometry for the scans when the \cxs is close to zero; panel B is a similar case, but only a few transits are flagged as blended; for comparison, panel C shows an example of a source that is always isolated and its \cxs is always close to zero; panel D however is a case of a source that was estimated as never crowded (\texttt{nBlend=0}) but shows clear variation with the scan angle. In this last case, the blending source was probably not in the source catalogue used by the crowding evaluation algorithm which was based on the \GDR2 source catalogue.
Note that in the crowded cases, the difference in scan angles between the peaks is about 180 degrees. This is because the positions of the sources with respect to the window (except for a mirror effect) will remain the same if all that is changed is the reversing of the scan direction.

Crowding effects are not the only reason for the \cxs variations. Different causes could be instrumental effects, calibration issues, cosmic rays. The epoch \cxs can then be used as a quality indicator in a statistical way. However we warn the users that filtering out transits on the base of a bad \cxs could hinder the study of special cases such as binaries or variables (see also Sec. 3.1 of \citealt{DR3-DPACP-173})). 




\subsection{Spurious periodicity}
Spurious periods found in the variability analysis can arise from different origins. Firstly, any noisy data, just by a random process, can mimic some signals that, in reality, are not there. In other words, we can generalize this idea by saying: there are always false positive detections in any statistical selection. 

Secondly, calibration residuals or the data acquisition strategy can leave their signatures in the calibrated data, so that one can take this signal for a true variability of the source. For example, we can find spurious signals and wrong periods (46 and 96 days) in the \Gaia\ astrometric and photometric data emerging from the scan-angle direction of extended or crowded sources \citep{DR3-DPACP-164}. The periods are consequences of \Gaia's scanning law.

Thirdly, the celestial source has a genuine signal, but the data analysis confuses it with a spurious period. Typical examples in sparsely sampled time series give rise to aliasing. 
The convolution theorem states that the Fourier transform of a product $f(t)*h(t)$ is the convolution of the individual Fourier transforms $F(\nu) \circledast H(\nu)$.
Therefore, the discrete Fourier transform  of the observations results from the convolution of the Fourier transform of the signal with the spectral window. This will then reflect the regularity pattern of the observing times.
If the signal has a simple low frequency, such as a trend or a long-term periodic phenomena, then the discrete Fourier transform will mainly reproduce the spectral window. It is thus possible that one of the peaks from this spectral window may have the highest amplitude within the searched frequency interval. 
In \Gaia, spurious frequencies at 4, 8, 12 cycles per day will be possible (see Appendix of \citealt{EyerEtal2017} for the spectral window structure).


\subsection{Reminders about features noted in \cite{EDR3Phot}}\label{Sect:RielloReminders}
The following is a list of known issues with the \EDR3 mean photometry discussed in Section 8 of \cite{EDR3Phot} with a comment on how they affect the epoch photometry in this survey.

\begin{itemize}
    \item Overestimated mean \BP\ flux for faint red sources\\
    This effect was caused by the filtering out of fluxes smaller than 1 e$^{-}$/s when forming the mean photometry. For the epoch photometry, no such filter was applied, however in the processing of the data for the archive, negative fluxes were excluded and do not appear in the survey. Note that low fluxes will be problematic if transformed to magnitudes. Plotting a colour magnitude diagram of all the epochs will demonstrate this.
    \item Sources with poor SSCs\\
    Of the 5\,401\,215 sources that were identified as having poor colour information in \EDR3, 1\,250 are within the area covered by GAPS. These sources do not have any mean \G\ photometry in the main section of the archive. For a more thorough explanation please go to the Known Issues web page for \EDR3. These sources do have epoch photometry in this survey, but it is very unreliable since they have been processed with the unreliable SSC values. 
    \item Systematics due to use of default colour in the Image  Parameter Determination (IPD)\\
    The systematic described in Section 8.3 of \cite{EDR3Phot} is not present in the mean \G\ photometry of the sources in the main archive nor in the epoch \G\ values within this survey since the correction described in this reference has already been applied.
    \item G–band magnitude term for blue and bright sources\\
    Eleven sources within this survey with $G<13$ and \BP$-$\RP$<-0.1$ are affected by the magnitude term in \G\ caused by this effect. This is probably caused by issues linked to the PSF/LSF calibration \citep{EDR3PSF}.
\end{itemize}

\section{Simple examples of data usage}\label{Sect:Examples}

\subsection{Correlations between the passbands}\label{Sect:Correlations}

An interesting way to detect variables with this data set is to look at the correlations between the three passbands in the residuals with respect to the mean for each source by use of Principal Component Analysis (PCA, \citealt{PCA}). See \cite{MariaPaper} for a similar investigation using SDSS data. For the \Gaia\ data, the analysis is limited to only 3 passbands, but this is sufficient for identifying variability, avoiding interference from systematic effects. This is based on the assumption that no correlation will exist between the passbands due to instrumental effects.
This shows how to exploit one of the most valuable features of this data set: simultaneous observations in many passbands.

The approach taken in this section is to generate for each epoch of a source three residuals with respect to the mean for that passband. These residuals are scaled using the estimated error for that residual, \ie quadrature addition of the mean source error, epoch error and the additional error described in Sect.~\ref{Sect:ErrorRenormalization}. An Eigen decomposition is carried out on these unit-weight residuals resulting in the principal components. The length of the first principal component (PC1) gives a strong indication of the variation within the photometric signals and the direction indicates how correlated the data between the passbands are. Another indication of correlation is the relative size of PC1 with respect to the other two principal components. The metric that was found to be most useful was $R=\nu_1/(\nu_2+\nu_3)$, where $\nu_n$ are the lengths of the Eigen vectors.

\begin{figure*}

  \resizebox{\hsize}{!}{\includegraphics{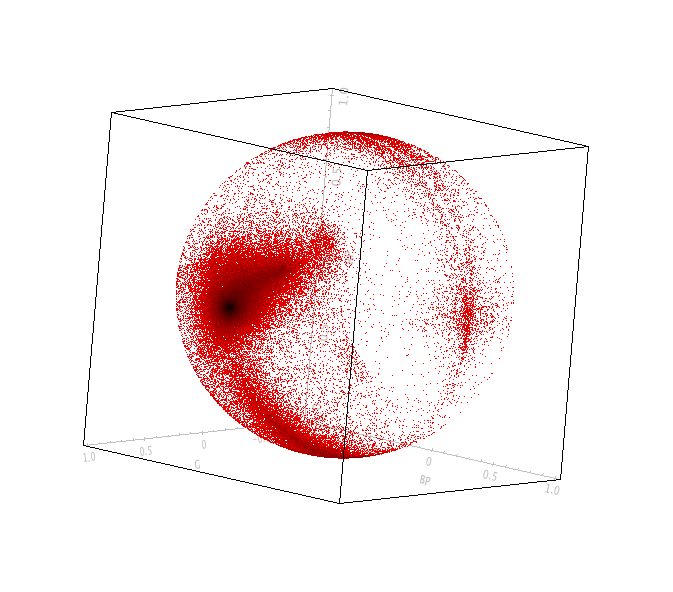}
  \includegraphics{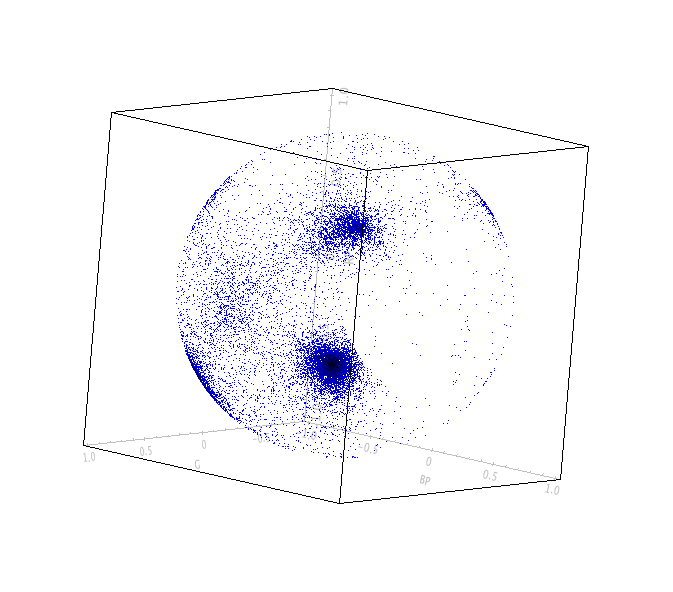}}
  \caption{The directions of the first Eigen vector (principal component) for all the sources in GAPS with more than 10 transits and $R>2.0$. The left plot are the results obtained simply scaling the input residuals by the errors, while in the one on the right the residuals are also scaled by the measured width of the distribution. The orientation of the axes is the same in both plots.}
  \label{Fig:EigenVectorDirections}
\end{figure*}

A problem with this method is that the errors for the \G\ passband are significantly smaller than those for \BP\ and \RP\ (see Fig.~\ref{Fig:ErrAll}). Since the amplitude of the variability in the three passbands are generally of the same order, scaling by the errors will mean that PC1 will generally be in the direction of the axis corresponding to the \G\ passband. This makes it difficult to distinguish between variability and a systematic solely in \G. This is shown in the left panel of Fig.~\ref{Fig:EigenVectorDirections}. Here the main concentration is towards \G, indicating that the most significant variation is in \G. The spur heading towards the diagonal (1, 1, 1) are the sources showing correlated variability.

To improve on this, the residuals are further scaled by the measured width of the distributions. This emphasizes when there is a strong correlation between the passbands and gives them equal weight. This is shown in the right-hand panel of Fig.~\ref{Fig:EigenVectorDirections}. The concentration of points in the \G\ direction are probably caused by uncalibrated systematics in the \G\ band. The two concentrations in the diagonal direction (1, 1, 1) are likely variables where there is a strong correlation between the passbands.

\begin{figure*}
  \resizebox{\hsize}{!}{\includegraphics{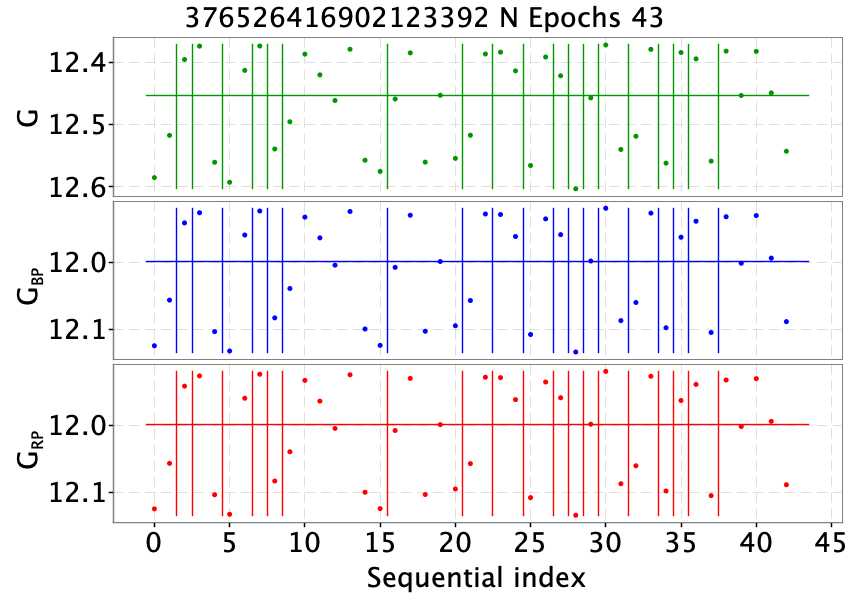}
  \includegraphics{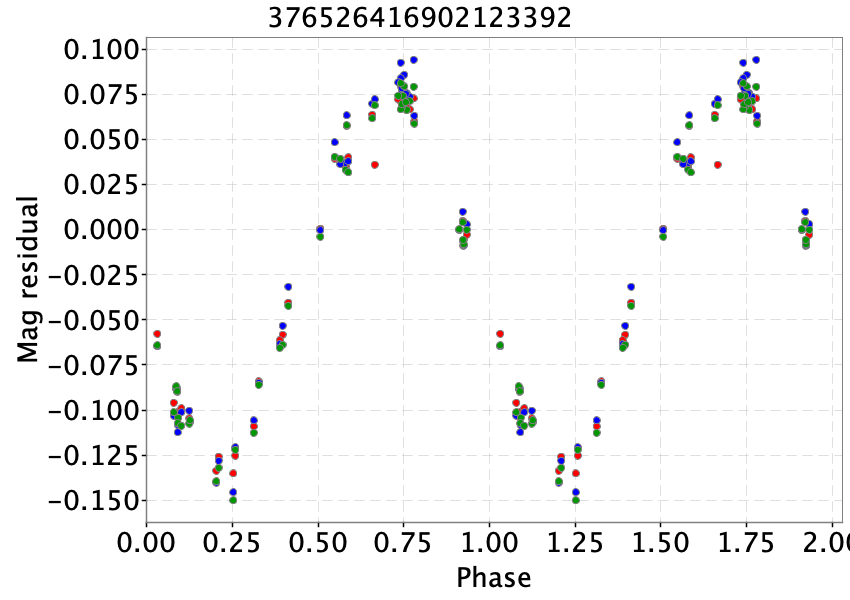}}
  \caption{An example of a variable identified using correlations between the three passbands. On the left the photometry is shown simply sorted by time. The vertical lines indicate a time gap of more than 0.5 day between two successive observations. The horizontal line is the median value.
  This is more useful than displaying as a function of time, due to the time sampling resulting from the scanning law of  \Gaia\ (see Fig.~\ref{Fig:TimeDistribution}). It is clear from this that the photometry is correlated. The plot on the right shows a folded light curve resulting from the period found.}
  \label{Fig:Variable}
\end{figure*}

Using selection limits of R$>$2 and a distance from the diagonal of less than 20$^{\circ}$, a selection of variables can be found using this method. Note that you should also select sources with sufficient transits. A limit of 10 FoV transits is suggested.
Figure \ref{Fig:Variable} shows an example of a periodic variable identified from this dataset that is not classified within the catalogue of variables released in \GDR3 (\citealt{DR3-DPACP-162} and \citealt{DR3-DPACP-165}). This star, 
\Gaia\ DR3 376526416902123392,
has already been identified as a variable in \cite{2018AJ....156..241H} and was classified as SINE. 

To identify the period used in Fig.~\ref{Fig:Variable},
we used the generalized Lomb–Scargle method (\citealt{ZechmeisterKurster}, \citealt{Lomb} and \citealt{Scargle}) and found a period of 0.214 days. 
Note that if you do a general period search on the GAPS data, you will find spurious concentrations at 0.0355 and 0.083 days.
These are caused by the scanning law or satellite rotation and are not due to variability. More detail on this can be found in \cite{DR3-DPACP-164}.

\subsection{HR Diagrams}
\begin{figure}
  \resizebox{\hsize}{!}{\includegraphics{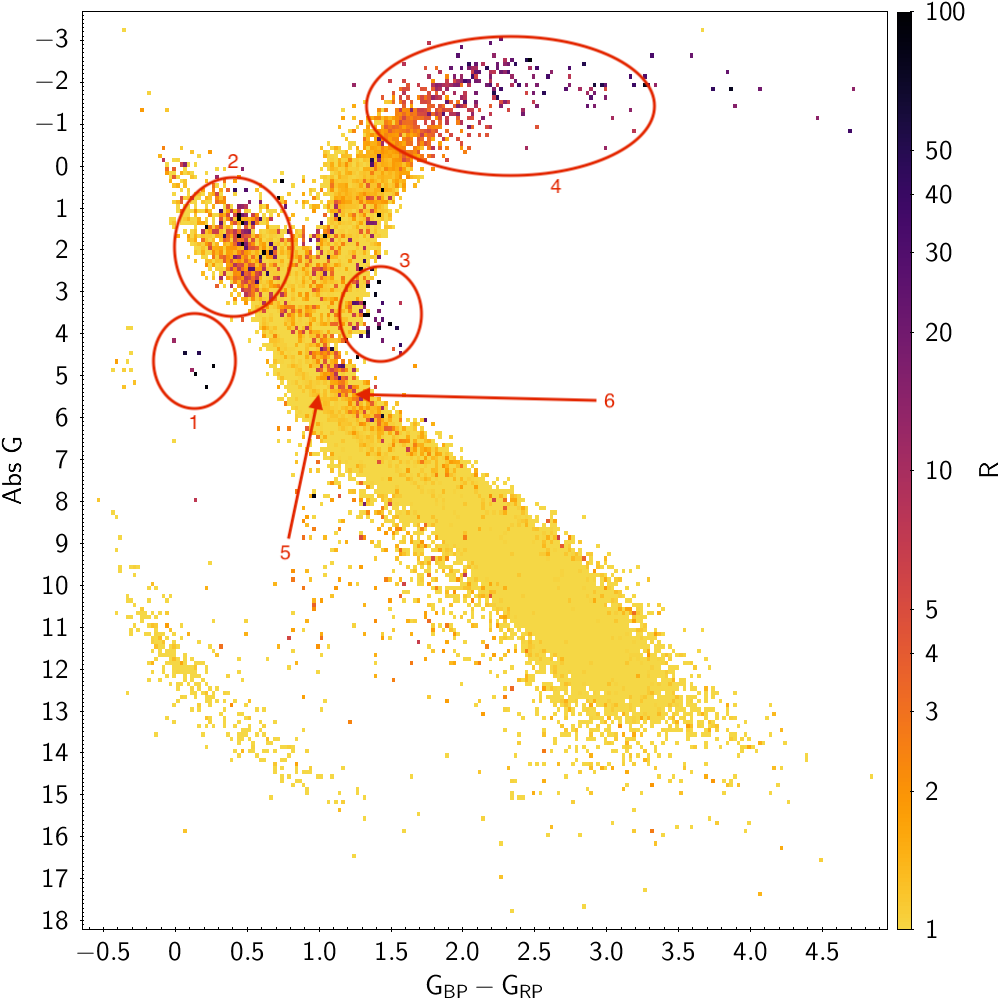}}
  \caption{Colour versus absolute magnitude with an auxiliary axis in R from Sect.~\ref{Sect:Correlations}. Only stars with better than a 10\% parallax error are used here. Rather than plotting each data point individually, the mean in an area is shown in this diagram. This is to avoid the overplotting of data points in the densest regions. The numbered regions are discussed in the text.}
  \label{Fig:HRD}
\end{figure}

Combining a HR diagram with the R correlation metric from Sect.~\ref{Sect:Correlations} provides an interesting and useful method of visualizing and identifying various variable stars. This is shown in Fig.~\ref{Fig:HRD}. Note that the stars plotted in this diagram are unlikely to be part of M31 due to the 10\% parallax error selection used.

Areas with high variability identified in Fig.~\ref{Fig:HRD} are
identified as follows:

\begin{enumerate}
    \item The seven very variable blue objects below the main sequence are Cataclysmic Variables. Four of these have been specifically identified as such by variability processing and
    are in \GDR3.
    \item This region shows the classical instability strip on the main sequence formed by $\delta$ Scuti stars  (p-mode pulsating stars from the $\kappa$ mechanism) and $\gamma$ Doradus stars (g-mode pulsating stars) at the lower luminosity. 
    \item The interesting clump with strong variability to the edge of the giant branch are RS Canum Venaticorum type variables. The majority of these have similarly been identified and are in \GDR3.
    \item The very red giants are long period variable stars (Miras, Semi Regular, OSARGs).
    \item The faint ridge in the middle of main sequence is surprising, it should be noted that a similar feature was present also for the fraction of variable stars in Fig.~8 of \cite{EyerEtal2019}. 
    \item Stronger ridge at top of main sequence is composed of different types of variability, many linked with binaries.
\end{enumerate}

At the red end of the main sequence, it might be expected that more variability is seen there than is evident in this figure. Fig.~8 of \cite{EyerEtal2019} indicates that, in general, this part of the HR diagram contains many variables. A reason that the R values are not higher here is that this metric depends on a correlation between all three \Gaia\ passbands and that the variability is strong with respect to the uncertainties of the epoch photometry. The stars typically inhabiting this part of the HR diagram will be faint and very red. A consequence of this is that the \BP\ fluxes will be very faint and the variability will not be significant compared to the uncertainties.

Similarly, it might be expected from Fig.~10 of \cite{EyerEtal2019}, that ZZ Ceti variables might be visible in this figure. In this case, there are not enough white dwarfs in the survey for these variables to be noticeable.








\subsection{Assessment of ``variability proxies"}
While no general variability metric is provided in the \Gaia\ releases so far, users have been inventive in using the available per-source statistics to provide information on variability. These have commonly been called ``variability proxies". Two such examples can be found in \cite{Belokurov2017} and \cite{Mowlavi2021}. In essence, both are the same in being the fractional error on the mean photometry multiplied by the square root of the number of observations. This is simply a consequence of the error estimate containing a scatter component, see \cite{DR1PhotPrinciplesErratum} Eq.~3. For constant stars, this gives the best estimate for the error on the mean even in the case where epoch errors have been under or over estimated. For variable stars, with a large variation, this gives an estimate of the variability. By multiplying by the square root of the number of observations, the error on the mean is effectively converted into a scatter measurement.

The advantage that GAPS has is that with the epoch data, more reliable variability metrics can be compared to a variability proxy which can be used for the DR3 catalogue where there is no epoch data.

\begin{figure}
  \resizebox{\hsize}{!}{\includegraphics{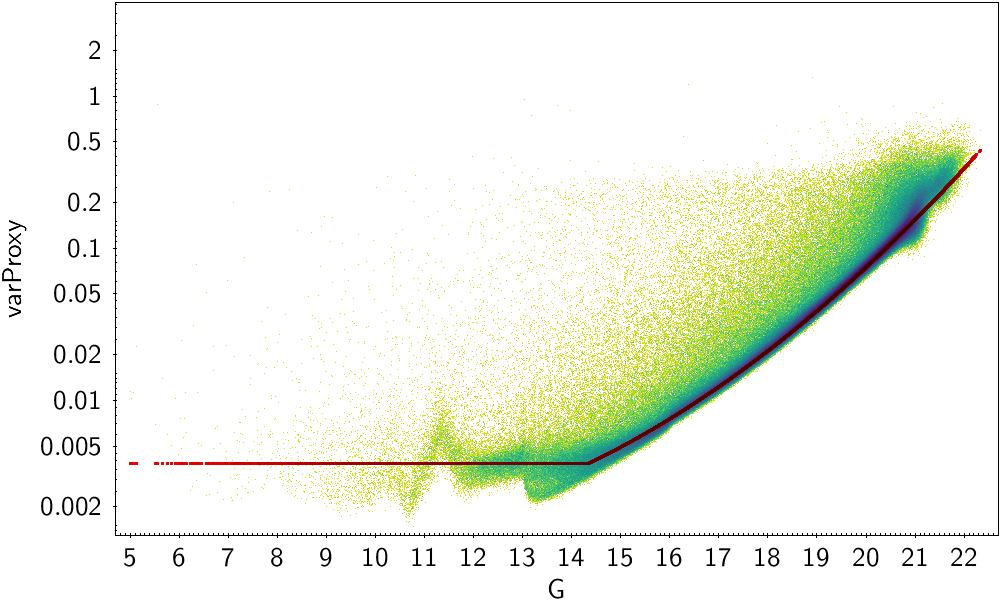}}
  \caption{The variability proxy from \cite{Mowlavi2021} versus \G\ magnitude along with a fit to the data described in the text by Eq.~\ref{Eq:varProxCorr}}
  \label{Fig:VarProxyFit}
\end{figure}

One idea to improve on the currently used variability proxies is to account for the intrinsic error of the photometry. This can be done by fitting the variability proxy as a function of magnitude and applying this correction to each value. This can be seen in Fig.~\ref{Fig:VarProxyFit}. In this case it is a simple quadratic fit in log space with a minimum limit \ie

\begin{equation}
\label{Eq:varProxCorr}
\log_{10} {\rm correction} = {\rm max}(-2.4134,  0.0127 G^2  -  0.2082 G  - 2.0423 )
\end{equation}

The initial idea was to subtract the correction from the variability proxy in quadrature. Of course, this will lead to taking a square root of a negative number in about half the cases, so leaving the metric without taking the square root is a better solution. This effectively makes the metric a corrected variance. In a similar vein, plotting the corrected variance in log space will remove about half the data points and will result in strange density distributions which naturally result from imperfections in the fit to derive the correction.

\begin{figure}
  \resizebox{\hsize}{!}{\includegraphics{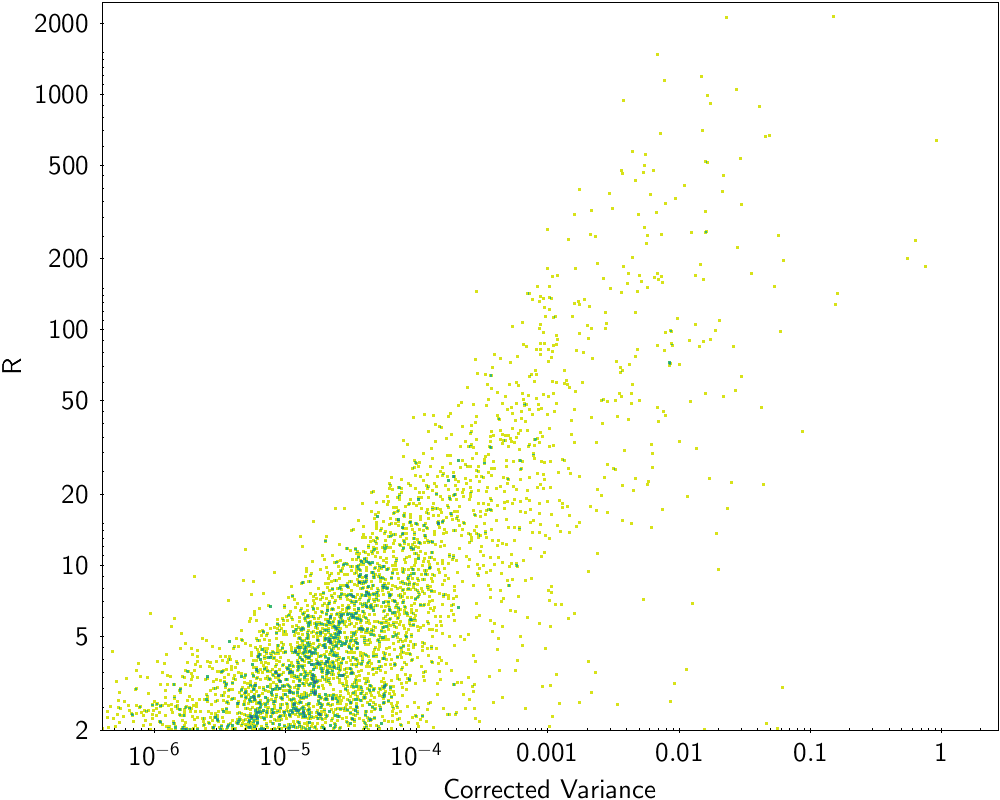}}
  \caption{Corrected variance versus R variability metric for sources with a distance from the diagonal of less than 20$^{\circ}$ and brighter than \G=14.}
  \label{Fig:VarProxyR}
\end{figure}

Figure~\ref{Fig:VarProxyR} shows the corrected variance, described above, plotted versus the R variability metric from Sect.~\ref{Sect:Correlations}. Selection limits of R$>$2, a distance from the diagonal of less than 20$^{\circ}$
and $G<14$ have been used in this plot. Not using the magnitude limit leaves many, probably spurious points, in the bottom right of the plot. This indicates that even after correction, the corrected variance (or any variability proxy) has difficulty in identifying variables unless they have large values. This is purely due to the difficulty in measuring scatters accurately enough as the sources get fainter.

\section{Conclusions}\label{Sect:Conclusions}


By delivering all the photometric data in this survey, the community is given an early opportunity to probe the quality of the \Gaia\ epoch photometry and see what level of variability detection can be achieved.
It is possible that artefacts could be found in the data that were not identified by the processing team. In this way, the community can participate in the ongoing iterative process which is improving the general quality of the \Gaia\ data. Note that some issues are known to the processing team but due to the time constraints of the complex DPAC processing schedule it is not always possible to address them all in time for the DR.

Also presented were alternative approaches on how to handle the three passbands using PCA, showing its usefulness.
Again, this survey will allow the community to develop alternative approaches to this
multivariate dataset.

The variance level as function of magnitude was calibrated so that intrinsic variability metrics can be derived, in order to help with the selection of variables from the mean photometry. Similarly, this survey can be used to estimate and establish the selection function of variability detection and help derive the expected number of true variables.

The community, by being exposed to this epoch photometry, can thus prepare itself for the future DR4 and DR5 data releases, where the photometric time series of all sources will be released.

Enjoy!

\begin{acknowledgements}
      Thanks to Michael Davidson, Claus Fabricius and Jordi Portell for interesting discussions on the effect of the major planets on the GAPS photometry due to higher background. (There is no effect!).
      
      Almost all the figures in this paper were generated using TOPCAT written by Mark Taylor.

      This work presents results from the European Space Agency (ESA) space mission \Gaia. \Gaia\ data are being processed by the \Gaia\ Data Processing and Analysis Consortium (DPAC). Funding for the DPAC is provided by national institutions, in particular the institutions participating in the \Gaia\ MultiLateral Agreement (MLA). The \Gaia\ mission website is \url{https://www.cosmos.esa.int/gaia}. The \Gaia\ archive website is \url{https://archives.esac.esa.int/gaia}.
Funding agency acknowledgements are given in Appendix~\ref{App:acknowledgements}

\end{acknowledgements}

\bibliographystyle{aa} 
\bibliography{refs} 

\appendix

\section{Downloading epoch photometry data from the \GDR{3} archive}\label{App:Downloading}

The epoch photometry data can be obtained using the Datalink feature of the archive. Other types of data such as BP/RP spectra can be obtained in a similar manner \cite[see][for instructions]{DR3-DPACP-118}. A dedicated tutorial is available at \url{https://www.cosmos.esa.int/web/gaia-users/archive/datalink-products#datalink_jntb_get_all_prods}.

In this section we provide an example of how to download epoch photometry data using the Python programming language.
It assumes that you already have a tailored list of source IDs to be extracted. Sources that have GAPS light curve data can be identified from the main \GDR3 catalogue using the query \verb+in_andromeda_survey = 't'+.

There is currently a limit of 5000 Datalink objects within a single query. In the following example, the input list is split into chunks of that size (or less) to overcome this restriction. A new FITS file for each chunk is created in the current folder. Pre-existing files are overwritten.

\begin{lstlisting}[frame=single]
import sys
import astroquery
from astropy.table import Table
from astroquery.gaia import GaiaClass, Gaia
import numpy as np

print("astroquery version: ", astroquery.__version__)

g = GaiaClass(gaia_tap_server='https://gea.esac.esa.int/',
              gaia_data_server='https://gea.esac.esa.int/')

print("Logging in")
g.login()

# In this example, it is a CSV file with one of the columns headed 'source_id' containing the source IDs
print("Reading in full list of sources")
big_list = Table.read("gaps-result.csv")['source_id'].data

def chunks(lst, n):
    """Yield successive n-sized chunks from lst."""
    for i in range(0, len(lst), n):
        yield lst[i:i + n]

chunk_max_size = 5000
chunks = list(chunks(big_list, chunk_max_size))

print(f'Input list contains {len(big_list)} source IDs')
print(f'This list is split into {len(chunks)} chunks of <= {chunk_max_size} elements each')

def get_datalink(chunk):
    n = chunk[0]
    chunk = chunk[1]
    print("Starting to load chunk " + str(n) + " with " + str(len(chunk)) + " sources")
    
    output_file=f'EPOCH_PHOTOMETRY_{n:04}.fits'
    
    result = g.load_data(ids=chunk.tolist(), data_release='Gaia DR3',
                         data_structure='COMBINED',
                         retrieval_type='EPOCH_PHOTOMETRY', 
                         format='fits',
                         avoid_datatype_check=True)
    
    epoch_key = [key for key in result.keys() if 'epoch_photometry' in key.lower()][0] 
    data = result[epoch_key][0] 
    data.write(output_file, format='fits', overwrite=True) 

for n, ids in zip(range(len(chunks)), chunks):
    get_datalink([n, ids])
\end{lstlisting}

\section{Bitwise coding for other\_flags field}\label{App:OtherFlags}

This field contains information on the data used to compute the fluxes and their quality. It generally provides debugging information that may be safely ignored for most applications. The field is a collection of binary flags, whose values can be recovered by applying bit shifting or masking operations. Each band has different binary flags in different positions, as shown below. The bit numbering is as follows: least significant bit = 1 and most significant bit = 64.

\bigskip
G band:
\begin{description}
\item  [Bit 1] SM transit rejected by photometry processing.
\item  [Bit 2] AF1 transit rejected by photometry processing.
\item  [Bit 3] AF2 transit rejected by photometry processing.
\item  [Bit 4] AF3 transit rejected by photometry processing.
\item  [Bit 5] AF4 transit rejected by photometry processing.
\item  [Bit 6] AF5 transit rejected by photometry processing.
\item  [Bit 7] AF6 transit rejected by photometry processing.
\item  [Bit 8] AF7 transit rejected by photometry processing.
\item  [Bit 9] AF8 transit rejected by photometry processing.
\item  [Bit 10] AF9 transit rejected by photometry processing.
\item  [Bit 13] G band flux scatter larger than expected by photometry processing (all CCDs considered).
\item  [Bit 14] SM transit unavailable by photometry processing.
\item  [Bit 15] AF1 transit unavailable by photometry processing.
\item  [Bit 16] AF2 transit unavailable by photometry processing.
\item  [Bit 17] AF3 transit unavailable by photometry processing.
\item  [Bit 18] AF4 transit unavailable by photometry processing.
\item  [Bit 19] AF5 transit unavailable by photometry processing.
\item  [Bit 20] AF6 transit unavailable by photometry processing.
\item  [Bit 21] AF7 transit unavailable by photometry processing.
\item  [Bit 22] AF8 transit unavailable by photometry processing.
\item  [Bit 23] AF9 transit unavailable by photometry processing.
\end{description}

\bigskip
BP band:
\begin{description}
\item[Bit 11] BP transit rejected by photometry processing.
\item[Bit 24] BP transit photometry rejected by variability processing.
\end{description}

\bigskip
RP band:
\begin{description}
\item[Bit 12] RP transit rejected by photometry processing.
\item[Bit 25] RP transit photometry rejected by variability processing.
\end{description}

\section{Funding Agency Acknowledgements}\label{App:acknowledgements}

The \Gaia\ mission and data processing have financially been supported by, in alphabetical order by country:
\begin{itemize}
\item the Tenure Track Pilot Programme of the Croatian Science Foundation and the \'{E}cole Polytechnique F\'{e}d\'{e}rale de Lausanne and the project TTP-2018-07-1171 `Mining the Variable Sky', with the funds of the Croatian-Swiss Research Programme;
\item the Agenzia Spaziale Italiana (ASI) through contracts I/037/08/0, I/058/10/0, 2014-025-R.0, 2014-025-R.1.2015, and 2018-24-HH.0 to the Italian Istituto Nazionale di Astrofisica (INAF), contract 2014-049-R.0/1/2 to INAF for the Space Science Data Centre (SSDC, formerly known as the ASI Science Data Center, ASDC), contracts I/008/10/0, 2013/030/I.0, 2013-030-I.0.1-2015, and 2016-17-I.0 to the Aerospace Logistics Technology Engineering Company (ALTEC S.p.A.), INAF, and the Italian Ministry of Education, University, and Research (Ministero dell'Istruzione, dell'Universit\`{a} e della Ricerca) through the Premiale project `MIning The Cosmos Big Data and Innovative Italian Technology for Frontier Astrophysics and Cosmology' (MITiC);
\item the Swiss State Secretariat for Education, Research, and Innovation through the Swiss Activit\'{e}s Nationales Compl\'{e}mentaires;

\item the United Kingdom Particle Physics and Astronomy Research Council (PPARC), the United Kingdom Science and Technology Facilities Council (STFC), and the United Kingdom Space Agency (UKSA) through the following grants to  the University of Cambridge: 
ST/I000542/1, 
ST/K000756/1, 
ST/N000641/1, 
ST/S000089/1, 
ST/W002469/1, 
ST/X00158X/1 and 
PP/D006546/1.
\end{itemize}

\section{Acronyms used in the paper}\label{App:acronyms}

\begin{table}[hp]
    \caption{\Gaia-related and other acronyms used in this paper. The first occurrence of the acronym is noted.}
    \label{Tab:acronyms}
    \centering
    \begin{tabular}{l|p{0.5\linewidth}|l}\hline\hline
Acronym & Description & See \\\hline

AC & ACross scan direction & Sect.~\ref{Sect:Data} \\
AF & Astrometric Field & Sect.~\ref{Sect:Stats} \\
BP & Blue Photometer & Sect.~\ref{Sect:Data} \\
CCD(s) & Charge Coupled Device(s) & Sect.~\ref{Sect:Data} \\
EPSL & Ecliptic Pole Scanning Law & Sect.~\ref{Sect:FieldChoice}\\
ESA & European Space Agency & Sect.~\ref{Sect:Introduction} \\
FoV(s) & Field(s) of View & Sect.~\ref{Sect:Data} \\
\EDR3 & \Gaia\ Early Data Release 3 & Sect.~\ref{Sect:Data} \\
\GDR3 & \Gaia\ Data Release 3 & Abstract \\
GAPS & \Gaia\ Andromeda Photometric Survey & Sect.~\ref{Sect:Introduction} \\
HR & Hertzsprung–Russell & Sect.~\ref{Sect:FieldChoice}\\
IPD & Image Parameter Determination & Sect.~\ref{Sect:RielloReminders} \\
LSF & Line Spread Function & Sect.~\ref{Sect:RielloReminders} \\
NSL & Nominal Scanning Law & Sect.~\ref{Sect:FieldChoice}\\
OBMT & On-Board Mission Time & Sect.~\ref{Sect:Data} \\
PCA & Principal Component Analysis & Sect.~\ref{Sect:Correlations} \\
PSF & Point Spread Function & Sect.~\ref{Sect:RielloReminders} \\
RP & Red Photometer & Sect.~\ref{Sect:Data} \\
SSC & Spectrum Shape Coefficient & Sect.~\ref{Sect:RielloReminders} \\
TCB & Barycentric Coordinate Time & Sect.~\ref{Sect:MagTerms} \\

\end{tabular}
\end{table}

\end{document}